\documentclass[fleqn,3p,onecolumn,11pt,nofootinbib,showkeys,showpacs]{revtex4-1}

\usepackage[small,justification=centerlast]{caption}
\usepackage{graphicx}
\usepackage{bm}
\usepackage{mathrsfs}
\usepackage[latin1]{inputenc}
\usepackage{array}
\usepackage{psfrag}
\usepackage{dsfont}
\usepackage{epsfig}
\usepackage{titletoc}
\usepackage{float}   
\usepackage{wrapfig} 
\usepackage{amsmath}\usepackage{amssymb,stmaryrd}
\usepackage{array}
\usepackage{epsfig}
\usepackage{cancel}
\usepackage[dvips]{feynmp}
\usepackage{upgreek}
\usepackage{subfig}
\usepackage{tabularx}
\usepackage{xcolor}
\usepackage{units}
\usepackage{textcomp}
\usepackage{wasysym}
\usepackage{amsthm}

\def\ring#1{{\mathaccent'27 #1}}

\usepackage{textfit} 

\usepackage{hyperref}

\renewcommand{\eqref}[1]{\mbox{Eq.~(\ref{#1})}}

\newcommand{\figref}[1]{\mbox{Fig.~\ref{#1}}}
\newcommand{\secref}[1]{\mbox{Sec.~\ref{#1}}}

\newcommand{\appref}[1]{\mbox{App.~\ref{#1}}}

\begin{document}

\title{Lorentz-violating modification of Dirac theory \\ based on spin-nondegenerate operators}

\author{J.A.A.S. Reis} \email{jalfieres@gmail.com}
\affiliation{Departamento de F\'{i}sica, Universidade Federal do Maranh\~{a}o \\
65080-805, S\~{a}o Lu\'{i}s, Maranh\~{a}o, Brazil}
\author{M. Schreck} \email{marco.schreck@ufma.br}
\affiliation{Departamento de F\'{i}sica, Universidade Federal do Maranh\~{a}o \\
65080-805, S\~{a}o Lu\'{i}s, Maranh\~{a}o, Brazil}

\begin{abstract}
The Standard-Model Extension (SME) parameterizes all possible Lorentz-violating contributions to the Standard Model and General Relativity.
It can be considered as an effective framework to describe possible quantum-gravity effects for energies much below the Planck energy.
In the current paper, the spin-nondegenerate operators of the SME fermion sector are the focus. The propagators, energies, and solutions
to the modified Dirac equation are obtained for several families of coefficients including nonminimal ones. The particle energies and
spinors are computed at first order in Lorentz violation and, with the optical theorem, they are shown to be consistent with the propagators.
The optical theorem is then also used to derive the matrices formed from a spinor and its Dirac conjugate at all orders in Lorentz violation.
The results are the first explicit ones derived for the spin-nondegenerate operators. They will prove helpful for future phenomenological
calculations in the SME that rely on the footing of quantum field theory.
\end{abstract}

\maketitle

\newpage
\setcounter{equation}{0}
\setcounter{section}{0}
\renewcommand{\theequation}{\arabic{section}.\arabic{equation}}

\section{Introduction}

Quantum-gravity effects may induce minuscule violations of Lorentz invariance. This is motivated by a number of articles that have been published
during the past 25 years. Lorentz symmetry violation was shown to occur in string-theory models \cite{Kostelecky:1988zi,Kostelecky:1989jp,Kostelecky:1989jw,Kostelecky:1991ak,Kostelecky:1994rn},
loop quantum gravity \cite{Gambini:1998it,Bojowald:2004bb}, noncommutative theories \cite{AmelinoCamelia:1999pm,Carroll:2001ws}, models
describing a small-scale structure of spacetime \cite{Klinkhamer:2003ec,Bernadotte:2006ya,Hossenfelder:2014hha}, quantum field theories
on spacetimes with nontrivial topology \cite{Klinkhamer:1998fa,Klinkhamer:1999zh}, and last but not least, Ho\v{r}ava-Lifshitz gravity
\cite{Horava:2009uw}.

Although there are specific models in various approaches to quantum gravity, it is highly challenging to extract generic physical statements
or principles from such models. After all, theoretical observations are very specific to a particular model under consideration and it is not clear at all whether
the same or similar results will be obtained within a different model. Understanding the physics of model after model always requires starting a
calculation from scratch and studying more general models may turn out to be impractical. Besides, it is not clear at all where
to look for Lorentz violation specifically, insofar there is no preference of any particle sector or any prototype of quantum gravity.

For these reasons, it is much more reasonable to have a general, effective framework available including all possible Lorentz-violating terms that are consistent
with coordinate invariance and the gauge structure of the Standard Model of elementary particles. Such a framework is provided by the
Standard-Model Extension (SME), which is a collection of all Lorentz-violating contributions in both the Standard Model and General Relativity \cite{Colladay:1996iz,Colladay:1998fq,Kostelecky:2003fs}.
Each Lorentz-violating term is decomposed into field operators and component coefficients that can be interpreted as background fields permeating
the vacuum. The SME allows for obtaining experimental predictions that are independent of any specific underlying model. The big advantage is that
large regions of coefficient space can be tested with a single experiment only, i.e., in principle many distinct models are covered by doing so.
Since every particle sector is contained, a large variety of experiments can be considered ranging from precise measurements of hyperfine
splitting in hydrogen to observations of ultra-high energy cosmic rays. In the physics community, there seems to be a greater interest in searches
for {\em CPT} violation than for Lorentz violation. In this context, one has to remark that {\em CPT} violation implies Lorentz violation
\cite{Greenberg:2002uu}, which is why all {\em CPT}-violating operators are contained in the SME automatically.

Lorentz-violating operators are classified according to their mass dimension. The finite number of operators with mass dimensions of 3 and 4 are
contained in the minimal SME \cite{Colladay:1998fq}. The remaining infinite number of higher-dimensional operators are part of the nonminimal SME
\cite{Kostelecky:2009zp,Kostelecky:2011gq,Kostelecky:2013rta}. Note that the nonlinear nonminimal gravity sector was discussed in
\cite{Bailey:2014bta}, where the recent paper \cite{Kostelecky:2016kfm} contains a classification of the nonminimal coefficients in the linearized
gravity sector. The nonminimal SME is a natural generalization of the minimal framework. It must
be kept in mind that nonminimal operators become more dominant for increasing energies. Also, due to the negative mass dimension of the commutativity
tensor in noncommutative field theories, such models can only be mapped to coefficients of the nonminimal SME, which justifies its consideration.

With the construction of the (minimal) SME, searches for violations of Lorentz invariance in nature have had their revival. This has led to a
steadily increasing number of cutting-edge experiments testing Lorentz invariance, which enlarges the set of constraints on Lorentz violation yearly.
At the same time, the sensitivity for detecting Lorentz invariance has been augmenting at a fast pace, leading to an improvement of constraints by several
orders of magnitude within few years, such as in the neutrino sector \cite{Kostelecky:2008ts}.

The theoretical properties of the SME at tree-level were investigated in a large series of papers \cite{Kostelecky:2000mm,oai:arXiv.org:hep-ph/0101087,Ferreira:2006kg,Casana:2009xs,Casana:2010nd,Klinkhamer:2010zs,Schreck:2011ai,Casana:2011fe,Hohensee:2012dt,Cambiaso:2012vb,Schreck:2013gma,
Schreck:2013kja,Schreck:2014qka,Maniatis:2014xja,Colladay:2014dua,Casana:2014cqa,Albayrak:2015ewa}, while radiative corrections were studied in
\cite{Jackiw:1999yp,Chung:1999pt,PerezVictoria:1999uh,PerezVictoria:2001ej,Kostelecky:2001jc,
Altschul:2003ce,Altschul:2004gs,Colladay:2006rk,Colladay:2007aj,Colladay:2009rb,Gomes:2009ch,Ferrero:2011yu,Casana:2013nfx,Scarpelli:2013eya,Cambiaso:2014eba,Santos:2014lfa,
Santos:2015koa,Borges:2016uwl,Belich:2016pzc}. The authors of \cite{Kostelecky:2000mm} examined properties of the minimal fermion sector
in general, whereas \cite{Ferreira:2006kg} is dedicated to the minimal $a$ and $b$ coefficients. Furthermore, in \cite{Schreck:2014qka} the
(nonminimal) fermion operators that are degenerate with respect to particle spin are on the focus. This concerns the $a$, $c$, $f$, and $m$ coefficients.
So far, the modified propagators and particle spinors have not been stated explicitly in the literature for the spin-nondegenerate cases, which
involves the $b$, $d$, $H$, and $g$ coefficients.\footnote{Note that even for these coefficients there are particular choices that have
a single dispersion relation for particles and antiparticles, respectively. Currently such a framework is under consideration and the
outcomes will be reported in a forthcoming paper.} The goal of the current paper is to fill this gap. Since all fields will be defined in
Minkowski spacetime, it suffices to consider explicit Lorentz symmetry breaking. Hence, the Lorentz-violating background fields are
introduced by hand and they lack any dynamics. Note that in curved spacetimes, this procedure is not sufficient, but one either has to break
Lorentz symmetry spontaneously \cite{Bluhm:2008yt,Hernaski:2014jsa,Bluhm:2014oua,Maluf:2014dpa} or one must work in an extended geometrical framework.
Finsler geometry seems to be very promising in this context and it has been on the focus for a couple of years \cite{Kostelecky:2010hs,Kostelecky:2011qz,Kostelecky:2012ac,Colladay:2012rv,Silva:2013xba,Schreck:2014ama,Russell:2015gwa,Schreck:2014hga,Foster:2015yta,Schreck:2015seb,
Schreck:2016jqn,Colladay:2015wra,Schreck:2015dsa,Silva:2015ptj,Silva:2016qkj}.

The paper is organized as follows. In
\secref{sec:propagators}, we state the propagators for the operators that break spin degeneracy. The results are valid for all
possible choices of $b$, $d$, $H$, and $g$ coefficients, no matter whether they are part of the minimal or nonminimal SME. In \secref{sec:dispersion-relations-spinors},
both the dispersion relations and the solutions of the modified Dirac equation are given at first order in Lorentz violation for specific
choices of minimal and nonminimal coefficients. The method of obtaining these solutions, which was developed in \cite{Kostelecky:2013rta},
will be reviewed.  Section \ref{sec:spinor-matrices-optical-theorem} is dedicated to obtaining the matrices constructed from a spinor and
its Dirac conjugate. These objects are indispensable in phenomenological calculations within quantum field theory. In this context, the optical
theorem is employed as a tool to obtain the matrices and also to check consistency between the propagators and the first-order solutions
of the Dirac equation. It is well-known that additional time derivatives in Lorentz-violating theories lead to several issues both in the
minimal and the nonminimal SME. How these can be resolved for specific cases will be outlined in \secref{sec:additional-time-derivatives}.
Last but not least, the results are summarized and discussed in \secref{sec:conclusions}. For demonstration purposes, exact spinors for a
limited number of coefficients can be found in \appref{sec:exact-spinors}. For completeness, in \appref{sec:results-spin-degenerate} we
give the spinors and the propagator for the spin-degenerate operators. Besides, we will state a couple of specific spinor matrices in
\appref{sec:special-spinor-matrices}. Natural units are used with $\hbar=c=1$, unless otherwise stated. For typesetting purposes,
momentum components will often have lower indices, but these should always be understood as components of the contravariant
momentum, though.

\section{SME fermion sector and propagators}
\label{sec:propagators}
\setcounter{equation}{0}

The construction of the minimal SME fermion sector was initiated in \cite{Colladay:1996iz,Colladay:1998fq}. Ultimately, the
Lagrange density of both the minimal and the nonminimal sector is expressed as follows~\cite{Kostelecky:2013rta}:
\begin{equation}
\mathcal{L}=\frac{1}{2}\overline{\psi}\left(\gamma^{\mu}\mathrm{i}\partial_{\mu}-m_{\psi}\mathds{1}_4+\hat{\mathcal{Q}}\right)\psi+\text{H.c.}
\end{equation}
Here, $\psi$ is a Dirac spinor and $\overline{\psi}\equiv \psi^{\dagger}\gamma^0$ is its Dirac conjugate. The fermion mass is denoted as
$m_{\psi}$ to distinguish it from one of the Lorentz-violating operators. Furthermore, $\gamma^{\mu}$ are the standard Dirac matrices
obeying the Clifford algebra $\{\gamma_{\mu},\gamma_{\nu}\}=2\eta_{\mu\nu}\mathds{1}_4$, with the Minkowski metric $\eta_{\mu\nu}$ and
the identity matrix $\mathds{1}_4$ in spinor space. Lorentz-violating contributions are contained in $\hat{\mathcal{Q}}$, which is a
$4\times 4$ matrix in spinor space as well. All fields are defined in Minkowski spacetime with metric signature $(+,-,-,-)$. In the
nonminimal SME, $\hat{\mathcal{Q}}$ is an expansion in terms of derivatives $\partial_{\mu}$, in
position space, or momenta, $p_{\mu}=\mathrm{i}\partial_{\mu}$, in momentum space. In spinor space, $\hat{\mathcal{Q}}$ is decomposed into the
16 Dirac bilinears:
\begin{equation}
\label{eq:decomposition-operator}
\hat{\mathcal{Q}}=\hat{\mathcal{S}}+\mathrm{i}\hat{\mathcal{P}}\gamma_5+\hat{\mathcal{V}}^{\mu}\gamma_{\mu}+\hat{\mathcal{A}}^{\mu}\gamma_5\gamma_{\mu}+\frac{1}{2}\hat{\mathcal{T}}^{\mu\nu}\sigma_{\mu\nu}\,,
\end{equation}
with the chiral Dirac matrix $\gamma_5\equiv \mathrm{i}\gamma^0\gamma^1\gamma^2\gamma^3$ and the commutator $\sigma_{\mu\nu}$ of two
Dirac matrices: $\sigma_{\mu\nu}\equiv \mathrm{i}/2[\gamma_{\mu},\gamma_{\nu}]$. The scalar, pseudoscalar,
and vector operators $\hat{\mathcal{S}}$, $\hat{\mathcal{P}}$, and $\hat{\mathcal{V}}$ contain the $a$, $c$, $e$, $f$, $m$, and $m_5$ coefficients,
cf.~Eq.~(7) in \cite{Kostelecky:2013rta}.
The $m_5$ coefficients can be absorbed into the physical fields by a chiral transformation but they are usually stated for completeness. The remaining
ones were subject to studies in \cite{Schreck:2014qka}. In the latter reference, certain properties of quantum field theories,
based on these coefficients, were examined. Such frameworks do not break spin degeneracy, which means that the
characteristics of a particle do not depend on the direction of the spin projection along the quantization axis. Hence, there is only a single
dispersion relation and a single spinor for particles and antiparticles, respectively. The situation is different for the spin-nondegenerate
operators, which are the pseudovector $\hat{\mathcal{A}}^{\mu}$ and the two-tensor $\hat{\mathcal{T}}^{\mu\nu}$. Those contain the $b$, $d$, $H$,
and $g$ coefficients and their nonminimal versions are given by \cite{Kostelecky:2013rta}
\begin{subequations}
\label{eq:definition-operators}
\begin{align}
\label{eq:definition-pseudovector-twotensor}
\hat{\mathcal{A}}^{\mu}&=\hat{d}^{\mu}-\hat{b}^{\mu}\,,\quad \hat{\mathcal{T}}^{\mu\nu}=\hat{g}^{\mu\nu}-\hat{H}^{\mu\nu}\,, \\[2ex]
\hat{b}^{\mu}&=\sum_{d \text{ odd}} b^{(d)\mu\alpha_1\dots\alpha_{d-3}}p_{\alpha_1}\dots p_{\alpha_{d-3}}\,, \\[2ex]
\hat{d}^{\mu}&=\sum_{d \text{ even}} d^{(d)\mu\alpha_1\dots\alpha_{d-3}}p_{\alpha_1}\dots p_{\alpha_{d-3}}\,, \\[2ex]
\hat{H}^{\mu\nu}&=\sum_{d \text{ odd}} H^{(d)\mu\nu\alpha_1\dots\alpha_{d-3}}p_{\alpha_1}\dots p_{\alpha_{d-3}}\,, \\[2ex]
\hat{g}^{\mu\nu}&=\sum_{d \text{ even}} g^{(d)\mu\nu\alpha_1\dots\alpha_{d-3}}p_{\alpha_1}\dots p_{\alpha_{d-3}}\,,
\end{align}
\end{subequations}
with the superscript $(d)$ giving the mass dimension of each coefficient. Here $b^{(d)\mu\alpha_1\dots\alpha_{d-3}}$ etc. are
called the component or controlling coefficients, which can be interpreted as background fields in the vacuum. The latter equations give the definitions of
the Lorentz-violating operators that will be examined in the current paper. The mass dimensions of the minimal $b$ and $H$ coefficients are~1,
whereas the minimal $d$ and $g$ coefficients are dimensionless. For each term in the expansions, the number of additional momentum components
increases by 2, successively, which is why the mass dimension of the appropriate coefficients decreases by 2. Moreover, the
associated $b$, $d$, $H$, and $g$ coefficients break spin degeneracy. Therefore, there are two possible dispersion relations and spinors for
both particles and antiparticles. We will encounter this behavior in the course of the article.

Before solving the modified Dirac equation, we are interested in the propagators of the frameworks that rest on the operators of
\eqref{eq:definition-operators}.
In the context of quantum field theory, a propagator describes a virtual particle that is generated at one spacetime point and annihilated at another one.
Propagators play a role in interaction processes where virtual particles occur. Thereby, due to the physical propagator poles, the contribution to the
probability amplitude becomes large whenever the momentum of any virtual particle is nearly on-shell. The propagator $\mathrm{i}S$ is the Green's function of the
free-field equations in momentum space, i.e., it is the inverse of the Dirac operator, $S^{-1}$, that appears in the field equations (multiplied by
an additional factor of $\mathrm{i}$ according to the conventions of \cite{Peskin:1995}). In momentum space
the latter is given by Eq.~(4) of \cite{Kostelecky:2013rta}:
\begin{equation}
\label{eq:dirac-operator}
S^{-1}=\gamma^{\mu}p_{\mu}-m_{\psi}\mathds{1}_4+\hat{\mathcal{Q}}\,.
\end{equation}
The Dirac operator is a $4\times 4$ matrix in spinor space. Thus, to obtain the propagator this matrix has to be inverted, for which the 16
Dirac bilinears $\{\Gamma^A\}\equiv\{\mathds{1}_4,\gamma_5,\gamma^{\mu},\mathrm{i}\gamma_5\gamma^{\mu},\sigma^{\mu\nu}\}$ are indispensable. Any complex $4\times 4$
matrix can be expanded in terms of $\{\Gamma_A\}$. Hence, for the inverse of the Dirac operator we propose the \textit{Ansatz}
\begin{equation}
\label{eq:propagator-ansatz}
\mathrm{i}S=\frac{\mathrm{i}}{\Delta}\left(\widehat{\xi}_{\mu}\gamma^{\mu}+\widehat{\Xi}\mathds{1}_4+\widehat{\Upsilon}\gamma^5+\widehat{\zeta}_{\mu}\gamma^5\gamma^{\mu}+\widehat{\psi}_{\mu\nu}\sigma^{\mu\nu}\right)\,,
\end{equation}
where $\Delta$ is the overall denominator of the propagator. There are now two possibilities of proceeding. First, \eqref{eq:propagator-ansatz} can be inserted
into $S^{-1}S=SS^{-1}=\mathds{1}_4$. This delivers a system of 16 linear equations in the 16 parameters $\{\widehat{\Xi},\widehat{\Upsilon},\widehat{\xi}_{\mu},\widehat{\zeta}_{\mu},\widehat{\psi}_{\mu\nu}\}/\Delta$
that are themselves functions of the four-momentum components, the particle mass, and the Lorentz-violating coefficients. It is no obstacle to solving this
system with computer algebra. The second possibility is to keep in mind that the Dirac bilinears obey the orthogonality relation
$\mathrm{Tr}(\Gamma_A\Gamma^B)=4\delta_A^{\phantom{A}B}$, where the appropriate Lorentz indices are lowered to obtain the dual basis $\{\Gamma_A\}$ . Multiplying
the inverse of the Dirac operator with each of the Dirac bilinears, each parameter can be obtained relying on the orthogonality condition.

\subsection{Propagator for $\boldsymbol{\hat{\mathcal{A}}^{\mu}}$}
\label{sec:propagator-pseudovector}

Now the propagator is obtained for the $b$ and $d$ coefficients, that are comprised in the observer Lorentz pseudovector $\hat{\mathcal{A}}^{\mu}$,
according to \eqref{eq:definition-pseudovector-twotensor}. The Dirac operator is given by \eqref{eq:dirac-operator}, where
$\hat{\mathcal{Q}}=\hat{\mathcal{A}}^{\mu}\gamma_5\gamma_{\mu}$. Either of the two procedures outlined above leads to the result
\begin{subequations}
\label{eq:propagator-pseudovector}
\begin{align}
\widehat{\Xi}_{\hat{\mathcal{A}}}&=m_{\psi}\left(p^2-m_{\psi}^2-\hat{\mathcal{A}}^2\right)\,, \displaybreak[0]\\[2ex]
\widehat{\Upsilon}_{\hat{\mathcal{A}}}&=0\,, \displaybreak[0]\\[2ex]
\widehat{\xi}^{\,\mu}_{\hat{\mathcal{A}}}&=(p^2-m_{\psi}^2+\hat{\mathcal{A}}^2)p^{\mu}-2(p\cdot\hat{\mathcal{A}})\hat{\mathcal{A}}^{\mu}\,, \displaybreak[0]\\[2ex]
\widehat{\zeta}^{\,\mu}_{\hat{\mathcal{A}}}&=2(p\cdot\hat{\mathcal{A}})p^{\mu}-(p^2+m_{\psi}^2+\hat{\mathcal{A}}^2)\hat{\mathcal{A}}^{\mu}\,, \displaybreak[0]\\[2ex]
\widehat{\psi}^{\,\mu\nu}_{\hat{\mathcal{A}}}&=m_{\psi}\varepsilon^{\mu\nu\varrho\sigma}p_{\varrho}\hat{\mathcal{A}}_{\sigma}\,, \displaybreak[0]\\[2ex]
\Delta_{\hat{\mathcal{A}}}&=(p+\hat{\mathcal{A}})^2(p-\hat{\mathcal{A}})^2-2m_{\psi}^2(p^2-\hat{\mathcal{A}}^2)+m_{\psi}^4\,,
\end{align}
\end{subequations}
with the four-dimensional totally antisymmetric Levi-Civita symbol, $\varepsilon^{\mu\nu\varrho\sigma}$, where $\varepsilon^{0123}=1$.
Several remarks are in order. First, the physical dispersion relations are poles of the propagator, i.e., they are zeros of the global denominator
$\Delta$. The result for $\Delta$ corresponds to the determinant of the Dirac operator that was obtained in Eq.~(4) of \cite{Schreck:2014ama}.
For component coefficients that are not associated with additional time derivatives, this is a fourth-order polynomial in $p_0$. Under that
condition, there are two particle dispersion relations and (after reinterpretation) two antiparticle dispersion relations.
Hence, the usual spin degeneracy of the particle energy is broken for this framework. Note that with additional time derivatives,
the number of poles even increases. Second, for vanishing Lorentz violation, $\hat{\mathcal{A}}^{\mu}=0$, the standard propagator
$\mathrm{i}S=\mathrm{i}(\cancel{p}+m_{\psi})/(p^2-m_{\psi}^2)$ is recovered. Third, the propagator, as it stands, is valid for both the minimal and nonminimal framework because
additional powers of the four-momentum, which are contained in $\hat{\mathcal{A}}^{\mu}$, do not modify the overall propagator structure.

\subsection{Propagator for $\boldsymbol{\hat{\mathcal{T}}^{\mu\nu}}$}
\label{sec:propagator-two-tensor}

The second framework considered is based on a nonvanishing observer two-tensor $\hat{\mathcal{T}}^{\mu\nu}$ in \eqref{eq:definition-pseudovector-twotensor},
which contains both the $H$ and the $g$ coefficients. The Dirac operator includes the modification $\hat{\mathcal{Q}}=\hat{\mathcal{T}}^{\mu\nu}\sigma_{\mu\nu}/2$.
The propagator can be obtained in analogy to the previous framework and its structure is a bit more involved:
\begin{subequations}
\label{eq:propagator-two-tensor}
\begin{align}
\widehat{\Xi}_{\hat{\mathcal{T}}}&=m_{\psi}(p^2-m_{\psi}^2+2X)\,, \displaybreak[0]\\[2ex]
\widehat{\Upsilon}_{\hat{\mathcal{T}}}&=-2\mathrm{i}m_{\psi}Y\,, \displaybreak[0]\\[2ex]
\widehat{\xi}^{\,\mu}_{\hat{\mathcal{T}}}&=(p^2-m_{\psi}^2-2X)p^{\mu}-2V^{\mu}\,, \displaybreak[0]\\[2ex]
\widehat{\zeta}^{\,\mu}_{\hat{\mathcal{T}}}&=2m_{\psi}\widetilde{\hat{\mathcal{T}}}{}^{\mu\nu}p_{\nu}\,, \displaybreak[0]\\[2ex]
\widehat{\psi}^{\,\mu\nu}_{\hat{\mathcal{T}}}&=\left[X-\frac{1}{2}(p^2+m_{\psi}^2)\right]\hat{\mathcal{T}}^{\mu\nu}+Y\widetilde{\hat{\mathcal{T}}}{}^{\mu\nu}+\hat{\mathcal{T}}^{\mu\varrho}p_{\varrho}p^{\nu}-p^{\mu}\hat{\mathcal{T}}^{\nu\varrho}p_{\varrho}\,, \displaybreak[0]\\[2ex]
\Delta_{\hat{\mathcal{T}}}&=(p^2-m_{\psi}^2-2X)^2+4(Y^2-V\cdot p-2m_{\psi}^2X)\,, \displaybreak[0]\\[2ex]
V^{\mu}&\equiv \hat{\mathcal{T}}^{\mu\nu}\hat{\mathcal{T}}_{\nu\varrho}p^{\varrho}\,, \displaybreak[0]\\[2ex]
X&\equiv\frac{1}{4}\hat{\mathcal{T}}^{\mu\nu}\hat{\mathcal{T}}_{\mu\nu}\,,\quad Y\equiv\frac{1}{4}\widetilde{\hat{\mathcal{T}}}{}^{\mu\nu}\mathcal{T}_{\mu\nu}\,,\quad \widetilde{\hat{\mathcal{T}}}{}^{\mu\nu}\equiv\frac{1}{2}\varepsilon^{\mu\nu\varrho\sigma}\hat{\mathcal{T}}_{\varrho\sigma}\,.
\end{align}
\end{subequations}
Again several remarks are in order. First, the zeros of the overall denominator $\Delta$ correspond to the physical dispersion relations.
Note that $\Delta$ is equal to the determinant of the Dirac operator, which is stated in Eq.~(15) of \cite{Schreck:2014ama}. Such as for
$\hat{\mathcal{A}}^{\mu}$, the denominator $\Delta$ has at least four zeros. Second, the propagator has a pseudoscalar part that is
characterized by a nonvanishing quantity, $Y$, linked to the dual tensor of $\hat{\mathcal{T}}^{\mu\nu}$.

It is interesting to observe that computing the
propagators is much easier than obtaining the spinors, as we shall see. The reason is that propagators are off-shell objects, where $p^0$ is
a free quantity. However, the spinor solutions of the Dirac equation are valid on-shell. Exact dispersion relations can become highly
complicated even for simple choices of component coefficients, for example when they produce third-order polynomials in $p^0$. It may
not even possible to give solutions of polynomials with order higher than four in a closed form. Hence, it is not surprising that the
complexity of the spinors follows the complexity of the dispersion laws.

\section{Dispersion relations and spinors}
\label{sec:dispersion-relations-spinors}
\setcounter{equation}{0}

The Lorentz-violating background field does not only modify the propagators but it changes the physical energy-momentum correspondences
as well. We have already observed this from the modified overall denominator in the propagator, whose poles correspond to these dispersion
relations on the one hand. On the other hand, it is possible to obtain the dispersion relations directly from the determinant of the Dirac
operator that changes due to Lorentz violation. The modified Dirac operator is stated in \eqref{eq:dirac-operator} with the Lorentz-violating
modification $\hat{\mathcal{Q}}$. In the current section, the Dirac equation will be solved for multiple
frameworks that are considered as exemplary. There are several procedures to do so.

The first method is to simply consider the Dirac equation in momentum space as a homogeneous linear system of equations whose solutions are the Dirac
spinors. For the spinor solutions to be nontrivial, the determinant of the Dirac operator has to vanish. This condition sets the hitherto arbitrary
zeroth four-momentum component to allowed energy levels. Since the Dirac operator is a $4\times 4$ matrix in spinor space, the polynomial in
$p^0$ is at least of degree 4. There are both positive and negative-energy values. The positive ones $E_>$ can be interpreted as the energies
for free particles, i.e., $p^0=E_>(\mathbf{p})$ in this case. The negative ones $E_<$ indicate the appearance of antiparticles, which are reinterpreted
according to the procedure of Feynman and St\"{u}ckelberg to give physically meaningful results. The reinterpretation works such that antiparticles
are considered as particles propagating backward in time, which leads to four-momentum components with opposite signs. Hence, for antiparticles
we have that $p^0=-E_<(-\mathbf{p})>0$ \cite{Kostelecky:2000mm,Kostelecky:2013rta,Schreck:2014qka}. With $p^0$ set to the allowed
energy values, the system of equations has nontrivial solutions for the spinors. For antiparticles, the Feynman-Stückelberg reinterpretation
has to be applied not only to the negative-energy solutions but to the whole Dirac operator. Solving the resulting system produces the spinors
for antiparticles.

The advantage of this general procedure is to deliver energies and spinors that are exact in Lorentz violation. However, the complexity of
both the energies and the spinors rises drastically the more controlling coefficients are taken into account. The reason
is that, even at second order in Lorentz violation, more and more possible products of distinct coefficients can be formed with an increasing
number of coefficients. For the purpose of demonstration two isotropic cases will be treated in \appref{sec:exact-spinors} according to the
method outlined.

It makes sense to obtain such exact solutions for theoretical reasons, e.g., to investigate the physical consistency of the framework under
consideration. However, for phenomenological studies it often suffices to restrain particle energies, spinors, etc. to first order in Lorentz
violation. The particle and antiparticle energies can then be obtained from the first-order Hamiltonians given by Eqs.~(60) and (64) of
\cite{Kostelecky:2013rta}, respectively. The method of computing the spinors is presented in Sec.~III.A of the latter paper. In this context, a
unitary matrix $U$ is constructed to block-diagonalize the Dirac operator, which decouples positive and negative-energy states. In the standard
case, the matrix $U$ is given by the product of two matrices, $V$ and $W$, with
\begin{equation}
\label{eq:transformation-matrices}
V=\frac{\mathds{1}_4+\gamma_0\gamma_5}{\sqrt{2}}\,,\quad W(\mathbf{p})=\frac{E_0+m_{\psi}+\mathbf{p}\cdot\boldsymbol{\gamma}}{\sqrt{2E_0(E_0+m_{\psi})}}\,,
\end{equation}
where $E_0=\sqrt{\mathbf{p}^2+m_{\psi}^2}$ is the standard dispersion relation for fermions \cite{Kostelecky:2013rta}. Note that for certain Lorentz-violating
frameworks such as those based on the $a$, $c$, $e$ and $m$ coefficients, the matrix $U$ that block-diagonalizes the Dirac
operator still has this form, where just the four-momentum or the particle mass is replaced by a suitable combination involving the
controlling coefficients (for details cf.~\appref{sec:results-spin-degenerate}). This works at all orders in Lorentz violation. However,
constructing $U$ for the $b$, $d$, $H$, and $g$ coefficients is much more involved, which is why the result is only known at first order
in Lorentz violation:
\begin{equation}
U=\left(\mathds{1}_4+\frac{1}{4E_0}[\gamma_5,VW\gamma_0\hat{\mathcal{Q}}W^{\dagger}V^{\dagger}]\right)VW\,,
\end{equation}
where $[\bullet,\bullet]$ denotes the ordinary commutator of two matrices \cite{Kostelecky:2013rta}. After block-diagonalization, the
Dirac equation is brought into the form $(E-H)U\psi=0$, with the energy $E$, the Hamiltonian $H$, and the spinor $\psi$. This is
an eigenvalue problem for the energies and the eigenvectors $U\psi$. Having these eigenvectors at hand allows for computing
the spinors by multiplying the eigenvectors with $U^{\dagger}$. The approach developed is similar to the Foldy-Wouthuysen procedure
\cite{Foldy:1949wa} in spirit. Note that all computations will be carried out with the Dirac matrices in the chiral representation:
\begin{subequations}
\label{eq:dirac-matrices-chiral-representation}
\begin{equation}
\gamma^0=\begin{pmatrix}
0 & \mathds{1}_2 \\
\mathds{1}_2 & 0 \\
\end{pmatrix}\,,\quad \gamma^i=\begin{pmatrix}
0 & \sigma^i \\
-\sigma^i & 0 \\
\end{pmatrix}\,,\quad \gamma_5=\begin{pmatrix}
-\mathds{1}_2 & 0 \\
0 & \mathds{1}_2 \\
\end{pmatrix}\,,
\end{equation}
using the Pauli matrices
\begin{equation}
\sigma^1=\begin{pmatrix}
0 & 1 \\
1 & 0 \\
\end{pmatrix}\,,\quad \sigma^2=\begin{pmatrix}
0 & -\mathrm{i} \\
\mathrm{i} & 0 \\
\end{pmatrix}\,,\quad \sigma^3=\begin{pmatrix}
1 & 0 \\
0 & -1 \\
\end{pmatrix}\,,
\end{equation}
\end{subequations}
and the identity matrix $\mathds{1}_2$ in two dimensions.

\subsection{General properties of particle and antiparticle energies}

In the standard case and for some Lorentz-violating frameworks such as those based on the $a$ and $c$ coefficients, the particle energy is
degenerate with respect to the particle spin, i.e., the energy is the same no matter in what direction along the quantization axis spin points.
However, the frameworks based on the $b$, $d$, $H$ and $g$ coefficients are nondegenerate with respect to the particle spin. Therefore, a
particle has two possible energy states that are denoted as $E_>^{(\pm)}\equiv E_u^{(\pm)}$, where the
ordering of the energies is chosen along the lines of App.~B in~\cite{Colladay:1996iz}. We use a notation for the particle energies
similar to that introduced in \cite{Kostelecky:2000mm}.

Since there exist two distinct positive energies, there are two distinct negative ones that will be called $E_<^{(\pm)}$. Based on
the transformation properties of controlling coefficients under charge conjugation $C$, the positive energies are related
to the negative ones as follows~\cite{Kostelecky:2000mm}:
\begin{equation}
E^{(\pm)}_{<}(\mathbf{p},d^{\mu\nu\dots},H^{\mu\nu\dots})=-E^{(\mp)}_{>}(-\mathbf{p},-d^{\mu\nu\dots},-H^{\mu\nu\dots})\,.
\end{equation}
Hence, only the $d$ and $H$ coefficients appear with opposite signs on both sides of the equation where the signs of the $b$ and $g$
coefficients remain unaffected. This is due to the fact that the $d$ and $H$ coefficients are odd under $C$, whereas the $b$ and $g$
coefficients are even (see Table P31 in \cite{Kostelecky:2008ts}).

The ordering of the energies is reversed as well according to App.~B of \cite{Colladay:1996iz} and
Eq.~(16) of \cite{Kostelecky:2000mm}. These negative energies are physically meaningless and they have to be reinterpreted according to
Feynman and St\"{u}ckelberg. By doing so, the signs of the spatial momentum components and of the energy, which corresponds to the
zeroth four-momentum component, have to be reversed:
\begin{equation}
-E^{(\mp)}_{>}(-\mathbf{p},-d^{\mu\nu\dots},-H^{\mu\nu\dots}) \mapsto E^{(\mp)}_{>}(\mathbf{p},-d^{\mu\nu\dots},-H^{\mu\nu\dots})\equiv E_v^{(\mp)}(\mathbf{p})\,,
\end{equation}
where we employ the notation of \cite{Kostelecky:2000mm} for the antiparticle energies as well.
Hence, ultimately the antiparticles have the latter positive energy values with both the signs of the $d$ and $H$ coefficients
reversed. As the $d$ and $H$ coefficients violate $C$, the particle energies are related to the antiparticle energies as follows:
\begin{equation}
E^{(\pm)}_u(\mathbf{p},d^{\mu\nu\dots},H^{\mu\nu\dots})=E^{(\mp)}_u(\mathbf{p},-d^{\mu\nu\dots},-H^{\mu\nu\dots})=E_v^{(\mp)}(\mathbf{p})\,.
\end{equation}

\subsection{Obtaining the spinors}
\label{sec:obtaining-spinors}

The great advantage of using the method of obtaining the spinors, which was outlined in \secref{sec:dispersion-relations-spinors}, will be
explained
as follows. First of all, it suffices to compute such a spinor for one particular nonzero component coefficient, such as for the minimal
coefficient $b^{(3)}_3$, defining a preferred direction pointing along the third spatial axis of the coordinate system. Having obtained
this result, allows for generalizing it to the pseudovector $\hat{\mathcal{A}}^{\mu}$ and
even to the appropriate nonminimal frameworks. For example, with the particle spinors for the coefficient $b^{(3)}_3$ at hand, we can
replace $b^{(3)}_3$ by $-(d^{(4)}_3-b^{(3)}_3)\equiv -\mathcal{A}^{(3)}_3$, cf.~Eq.~(7) in \cite{Kostelecky:2013rta}, which is the
generalization within the minimal SME. In a further step at first order in Lorentz violation, the latter $\mathcal{A}_3^{(3)}$ is
promoted to an operator, which corresponds to generalizing it to the nonminimal SME:
\begin{equation}
\mathcal{A}^{(3)3}\mapsto \hat{\mathcal{A}}^3\equiv \sum_{d \text{ odd}} \mathcal{A}^{(d)3\alpha_1\dots\alpha_{d-3}}p_{\alpha_1}\dots p_{\alpha_{d-3}}\,.
\end{equation}
Similarly, this works for the tensor coefficients $H$ and $g$. As a first step, it is reasonable to obtain the spinors for a particular nonzero $H$
coefficient, such as for $H^{(3)}_{01}$. The results are then generalized within the minimal SME according to Eq.~(7) of
\cite{Kostelecky:2013rta}, i.e., we replace $H^{(3)}_{01}$ by $-(g^{(4)}_{01}-H^{(3)}_{01})\equiv -\mathcal{T}^{(3)}_{01}$.
The generalization to the nonminimal SME amounts to promoting $\mathcal{T}^{(3)}_{01}$ to an operator:
\begin{equation}
\mathcal{T}^{(3)01}\mapsto \hat{\mathcal{T}}^{01}\equiv \sum_{d \text{ odd}} \mathcal{T}^{(d)01\alpha_1\dots\alpha_{d-3}}p_{\alpha_1}\dots p_{\alpha_{d-3}}\,.
\end{equation}
So it is wise to start with the simplest component coefficients in this context, which are the minimal $b$ and $H$ coefficients
and to generalize the spinors obtained based on the previous description. Note that for the spinors at first order in Lorentz violation
it is sufficient to replace all additional $p^0$ components in the Lorentz-violating terms of the particle and antiparticle spinors by
the standard energy $E_0$. It has to be kept in mind, though, that the simple generalization outlined
above only works at first order in Lorentz violation. The exact spinors are expected to involve complicated combinations
of distinct coefficients of both the minimal and nonminimal SME such as $b^{(3)}_3$ and~$d^{(6)}_{3111}$.

\subsection{Normalization}

The positive-energy spinors (for particles) are denoted as $u^{(\alpha)}$ and the negative-energy spinors (reinterpreted
for antiparticles according to Feynman and St\"{u}ckelberg) are denoted as $v^{(\alpha)}$. Additional subscripts $E^{(\pm)}_u$ will
indicate whether a spinor is associated to an energy $E_u^{(+)}$ or $E_u^{(-)}$. This distinction is necessary because of broken
spin degeneracy. For the latter reason, both the two particle spinors and the antiparticle spinors are
automatically orthogonal to each other. Furthermore, the spinors should be normalized. We choose a normalization such that the spinors
satisfy
\begin{subequations}
\label{eq:orthogonality-relations}
\begin{align}
u^{(\alpha)\dagger}u^{(\alpha)}&=2E_u^{(\alpha)}\,, \\[2ex]
v^{(\alpha)\dagger}v^{(\alpha)}&=2E_v^{(\alpha)}\,.
\end{align}
\end{subequations}
Note that $\alpha$ is not summed over on the left-hand sides. This normalization has a great advantage when dealing with the optical
theorem. There is then no additional global adjustment necessary in the optical theorem to make it work out. We will observe this in
\secref{sec:spinor-matrices-optical-theorem}. Except of a factor $2/m_{\psi}$ on the right-hand sides of these conditions, our normalization
corresponds to the normalization of spinors chosen for the minimal SME in~\cite{Kostelecky:2000mm}.

\subsection{Energies and spinors for $\boldsymbol{\hat{\mathcal{A}}^{\mu}}$}

We start computing the spinors for the pseudovector operator $\hat{\mathcal{A}}^{\mu}$. All results are understood to be valid for a positive
expansion parameter, i.e., for a positive combination of Lorentz-violating coefficients and four-momentum components. If this combination is
negative, the labels of the dispersion relations and the spinor solutions just have to be switched. We will elaborate on this at the very end of
\secref{sec:spinor-matrices-optical-theorem}.

\subsubsection{Full ``isotropic'' operator}
\label{sec:isotropic-operator}

It is always a good advice to start with an isotropic framework. First, this is due to practical reasons, since computations are expected to be much
simpler when there is no direction dependence. Second, from a phenomenological point of view, controlling coefficients that are associated with
preferred spacelike directions are often constrained more strictly in comparison to their isotropic counterparts.
The leading-order terms in the dimensional expansion of the $b$, $d$, and $g$ coefficients are chosen to be isotropic, which will be indicated by
the standard notation of a ring diacritic (cf.~Sec.~IV.B in \cite{Kostelecky:2013rta}). For the minimal $b$ coefficients, only the zeroth component
of the coefficient vector leads to an isotropic dispersion relation, i.e., $\ring{b}\equiv b^{(3)0}$. The minimal $d$ coefficient matrix has to
be chosen as a diagonal (traceless) matrix with the spatial components equal to each other, i.e.,
$d^{(4)\mu\nu}=\ring{d}\times\mathrm{diag}(1,1/3,1/3,1/3)^{\mu\nu}$, where $\ring{d}\equiv d^{(4)00}$. Last but not least, the minimal $g$
coefficients produce isotropic expressions if all spatial coefficients with a totally
antisymmetric permutation of indices are equal modulo the sign of the permutation: $g^{(4)ijk}=\ring{g}\times\varepsilon^{ijk}$, with $\ring{g}\equiv g^{(4)123}$.
There is no isotropic choice for the minimal $H$ coefficients. Respecting these leading-order choices, we define the total ``isotropic'' operators as
follows:
\begin{subequations}
\begin{align}
\hat{\ring{\mathcal{A}}}&\equiv \hat{\ring{d}}{}^0-\hat{\ring{b}}\,, \displaybreak[0]\\[2ex]
\hat{\ring{b}}&\equiv\sum_{d \text{ odd}} b^{(d)0\alpha_1\dots\alpha_{d-3}}p_{\alpha_1}\dots p_{\alpha_{d-3}}\,, \displaybreak[0]\\[2ex]
\hat{\ring{d}}{}^{\mu}&\equiv \hat{\ring{d}}\times\mathrm{diag}\left(1,\frac{1}{3},\frac{1}{3},\frac{1}{3}\right)^{\mu\alpha_1}p_{\alpha_1}\,,\quad \hat{\ring{d}}\equiv \sum_{d \text{ even}} d^{(d)00\alpha_2\dots\alpha_{d-3}}p_{\alpha_2}\dots p_{\alpha_{d-3}}\,, \displaybreak[0]\\[2ex]
\hat{\ring{g}}{}^{ij}&\equiv \hat{\ring{g}}\times \varepsilon^{ijk}p_k\,,\quad \hat{\ring{g}}\equiv\sum_{d \text{ even}} g^{(d)123\alpha_2\dots\alpha_{d-3}}p_{\alpha_2}\dots p_{d-3}\,.
\end{align}
\end{subequations}
Including the higher-dimensional contributions, these choices are certainly no longer isotropic. However, they will still be denoted with a ring
diacritic to remind the reader of the isotropic nature of the leading-order (minimal) terms. The modified dispersion relations, at first order in the
controlling coefficients, read
\begin{equation}
E_u^{(\pm)}|^{\bigcirc}=E_0\left[1\pm \frac{|\mathbf{p}|}{E_0^2}\left(\frac{4}{3}E_0\hat{\ring{d}}-\hat{\ring{b}}+m_{\psi}\hat{\ring{g}}\right)\right]\,.
\end{equation}
Relying on the second procedure reviewed at the beginning of the current section, the particle spinors at first order in Lorentz violation
are calculated and cast into the following shape:
\begin{subequations}
\begin{align}
\label{eq:particle-spinors-b0}
u^{(1,2)}|_{E_u^{(\pm)}}^{\bigcirc}&=\ring{N}^{(1,2)}_u\ring{U}\left(\hat{\ring{\mathcal{A}}}-\hat{\ring{g}}\frac{\mathbf{p}^2}{m_{\psi}}\right)\,, \\[2ex]
\ring{U}(X)&=\begin{pmatrix}
\boldsymbol{\ring{\phi}}_{\mp} \\
\boldsymbol{\ring{\chi}}_{\pm} \\
\end{pmatrix}-\frac{m_{\psi}}{2E_0^2}\begin{pmatrix}
\boldsymbol{\ring{\chi}}_{\pm} \\
-\boldsymbol{\ring{\phi}}_{\mp} \\
\end{pmatrix}X\,, \displaybreak[0]\\[2ex]
\boldsymbol{\ring{\phi}}_{\pm}&=[E_0+m_{\psi}\pm| \mathbf{p}|]\begin{pmatrix}
p_3\mp|\mathbf{p}| \\
p_1+\mathrm{i}p_2 \\
\end{pmatrix}\,, \displaybreak[0]\\[2ex]
\boldsymbol{\ring{\chi}}_{\pm}&=[E_0+m_{\psi}\pm |\mathbf{p}|]\begin{pmatrix}
p_3\pm|\mathbf{p}| \\
p_1+\mathrm{i}p_2 \\
\end{pmatrix}\,,
\end{align}
with the normalization factors
\begin{equation}
\ring{N}^{(1,2)}_u=\frac{1}{4\sqrt{|\mathbf{p}|(|\mathbf{p}|\pm p_3)(E_0+m_{\psi})}}\left[2\pm \frac{|\mathbf{p}|}{E_0^2}\left(\frac{4}{3}E_0\hat{\ring{d}}-\hat{\ring{b}}+m_{\psi}\hat{\ring{g}}\right)\right]\,.
\end{equation}
\end{subequations}
The isotropic spinors only depend on the magnitude of the momentum, as expected. The spinors contain a small number of functions dependent on the
energy, momentum, and mass with different combinations of signs. Thereby, the Lorentz-violating contribution does not introduce any terms with a
different structure. For vanishing Lorentz violation the spinors do not reproduce the standard results given in most textbooks. However,
two issues must be taken into account in this context. Firstly, in most textbooks, the spinors are given for the Dirac representation of the Dirac
matrices, whereas here the chiral representation is used, cf.~\eqref{eq:dirac-matrices-chiral-representation}. Secondly, the standard method to
solving the Dirac equation with a decomposition into two-component spinors does not work for the $b$, $d$, $H$, and $g$ coefficients. Therefore,
the structure of these spinors for vanishing Lorentz violation is more involved than the structure of the usual spinors obtained for the chiral
representation of Dirac matrices. Nevertheless, the more complicated standard spinors, deduced from \eqref{eq:particle-spinors-b0}, satisfy the
standard Dirac equation. Additionally, the Lorentz-violating term is suppressed by the ratio of the fermion mass and the standard particle energy
$E_0$. The combination of $m_{\psi}/E_0^2$ and the Lorentz-violating operators is dimensionless to make the second summand match the mass dimension
of the first.

Note that the widely used field-theory model by Myers and Pospelov \cite{Myers:2003fd} can be mapped to effective dimension-5 $a$ and dimension-6 $g$
coefficients, cf.~Eqs.~(27), (156), and (157) in \cite{Kostelecky:2013rta}. Therefore, it is possible to treat this particular model based on the
combined results of \cite{Kostelecky:2013rta,Schreck:2014qka} and the current article. Choosing a special observer frame, with a purely timelike
preferred direction $n^{\mu}=(1,0,0,0)^{\mu}$, allows for identifying the Myers-Pospelov Lagrangian with a subset of the general isotropic SME fermion
sector for $d=5$, 6, which is given by Eqs.~(97), (98) of \cite{Kostelecky:2013rta}. The model of \cite{Myers:2003fd} is a special case of the latter
equations, where the isotropic effective coefficients $\ring{a}_0^{(5)}$ and $\ring{g}_1^{(6)}$ are nonvanishing only. Hence, there is a single $d=5$
and a single $d=6$ isotropic degree of freedom out of the eight possible ones for dimension 5 and 6 (see Table III in \cite{Kostelecky:2013rta}).
The degree of freedom $\ring{g}_1^{(6)}$ is a combination of the nonvanishing component coefficients $b^{(5)000}\equiv b_5$ and
$g^{(6)ijk00}\equiv\varepsilon^{ijk}g_6$. Then, the above energies and spinors are valid with
\begin{equation}
\hat{\ring{b}}=b_5E_0^2\,,\quad \hat{\ring{d}}=0\,,\quad \hat{\ring{g}}=g_6E_0^2\,.
\end{equation}

\subsubsection{Anisotropic operator $\hat{\mathcal{A}}^i$ with $i=\{1,2\}$}
\label{sec:operator-ahat-12}

After establishing the isotropic result, we intend to consider anisotropic Lorentz violation in the realm of the pseudovector $\hat{\mathcal{A}}^{\mu}$.
The preferred direction is chosen to either point along the first or the second spatial axis of the coordinate system, as both cases can be treated in
one go. The particle spinors are obtained in the same manner as before, based on the perturbative method reviewed at the very beginning of the
current section. First of all, the particle dispersion relations at first order in Lorentz violation are given by
\begin{equation}
\label{eq:energy-pseudovector}
E_u^{(\pm)}|^{\hat{\mathcal{A}}^i}=E_0\left(1\pm\frac{S_i}{E_0^2}\hat{\mathcal{A}}^i\right)\,, \quad S_i=\sqrt{p_i^2+m_{\psi}^2}\,.
\end{equation}
Quantities such as $S_i$ are characteristic for anisotropic frameworks and they will appear at various places in the spinors. For all cases that
follow, we will only state the first particle spinor $u^{(1)}$ that is connected to the particle energy $E_u^{(+)}$. The remaining spinors can be
computed from $u^{(1)}$ by simple transformations, cf.~\secref{sec:remaining-spinors-pseudovector}. In this context the spinors for the components
$\hat{\mathcal{A}}^1$ and $\hat{\mathcal{A}}^2$ are closely related to each other, being expressed in terms of a master function $\breve{U}$:
\begin{subequations}
\label{eq:particle-spinors-b1-b2}
\begin{align}
\label{eq:particle-spinors-b1}
u^{(1)}|_{E_u^{(+)}}^{\hat{\mathcal{A}}^1}&=\breve{N}_u^{(1)}\breve{U}\left(\mathbf{p},\hat{\mathcal{A}}^1,S_1\right)\,, \displaybreak[0]\\[2ex]
\label{eq:particle-spinors-b2}
u^{(1)}|^{\hat{\mathcal{A}}^2}_{E_u^{(+)}}&=\breve{N}_u^{(1)}\begin{pmatrix}
-\mathrm{i}\breve{U}_1 \\
\breve{U}_2 \\
-\mathrm{i}\breve{U}_3 \\
\breve{U}_4 \\
\end{pmatrix}\left(p_1=p_2,p_2=-p_1,p_3,\hat{\mathcal{A}}^2,S_2\right)\,, \displaybreak[0]\\[2ex]
\breve{U}(\mathbf{p},X,S_i)&=\begin{pmatrix}
\boldsymbol{\breve{\phi}}_+ \\
\boldsymbol{\breve{\phi}}_- \\
\end{pmatrix}+\frac{1}{2E_0^2}\begin{pmatrix}
\delta\boldsymbol{\breve{\phi}}_+ \\
\delta\boldsymbol{\breve{\phi}}_- \\
\end{pmatrix}X\,, \displaybreak[0]\\[2ex]
\boldsymbol{\breve{\phi}}_{\pm}&=\begin{pmatrix}
\breve{A}_{\pm} \\
\breve{B}_{\pm} \\
\end{pmatrix}\,,\quad \delta\boldsymbol{\breve{\phi}}_{\pm}=\pm\mathrm{i}p_2\begin{pmatrix}
\breve{A}_{\pm} \\
-\breve{B}_{\pm} \\
\end{pmatrix}\pm p_3\begin{pmatrix}
-\breve{B}_{\pm} \\
\breve{A}_{\pm} \\
\end{pmatrix}\,, \displaybreak[0]\\[2ex]
\breve{A}_{\pm}&=(\pm p_3-E_0)(S_i\pm p_1)+m_{\psi}(\pm\mathrm{i}p_2-S_i)\,, \displaybreak[0]\\[2ex]
\breve{B}_{\pm}&=m_{\psi}(E_0+m_{\psi}\pm p_3)+(p_1+\mathrm{i}p_2)(p_1\pm S_i)\,,
\end{align}
with the normalization factor
\begin{equation}
\breve{N}_u^{(1)}(\mathbf{p},X,S_i)=\frac{1}{4\sqrt{(E_0+m_{\psi})S_i^2-p_1p_3S_i}}\left(2+\frac{S_i}{E_0^2}X\right)\,.
\end{equation}
\end{subequations}
The structure of the spinors is evidently a bit more involved in comparison to the ``isotropic'' case of \secref{sec:isotropic-operator}.
The spinors can still be expressed via two different functions including two sign choices for each. The transformations
$p_1=p_2$, $p_2=-p_1$ have to be applied to the normalization factor as well, when computing the spinor for $\hat{\mathcal{T}}^{02}$.
However, note that the momentum components contained in the Lorentz-violating operator itself stay unaffected.

\subsubsection{Anisotropic operator $\hat{\mathcal{A}}^3$}
\label{sec:anisotropic-pseudovector-a3}

For a framework with a nonzero $\hat{\mathcal{A}}^3$, the energies correspond to \eqref{eq:energy-pseudovector},
where just the Lorentz-violating operator and the quantity $S_3$ have to be adapted accordingly:
\begin{equation}
E_u^{(\pm)}|^{\hat{\mathcal{A}}^3}=E_0\left(1\pm\frac{S_3}{E_0^2}\hat{\mathcal{A}}^3\right)\,, \quad S_3=\sqrt{p_3^2+m_{\psi}^2}\,.
\end{equation}
The spinors for this framework can be expressed in a very convenient form:
\begin{subequations}
\begin{align}
\label{eq:particle-spinors-b3}
u^{(1)}|_{E_u^{(+)}}^{\hat{\mathcal{A}}^3}&=\bar{N}_u^{(1)}\bar{U}(\hat{\mathcal{A}}^3)\,, \displaybreak[0]\\[2ex]
\bar{U}(X)&=\begin{pmatrix}
\boldsymbol{\bar{\phi}}_+ \\
\boldsymbol{\bar{\phi}}_- \\
\end{pmatrix}+\frac{p_1+\mathrm{i}p_2}{2E_0^2}\begin{pmatrix}
\delta\boldsymbol{\bar{\phi}}_+ \\
\delta\boldsymbol{\bar{\phi}}_- \\
\end{pmatrix}X\,, \displaybreak[0]\\[2ex]
\boldsymbol{\bar{\phi}}_{\pm}&=\begin{pmatrix}
\bar{A}_{\pm} \\
\bar{B}_{\mp} \\
\end{pmatrix}\,,\quad
\label{eq:definition-matrix-epsilon}
\delta\boldsymbol{\bar{\phi}}_{\pm}=\pm\varepsilon\cdot\boldsymbol{\bar{\phi}}_{\pm}^{*}\,,\quad \varepsilon=\begin{pmatrix}
0 & 1 \\
-1 & 0 \\
\end{pmatrix}\,, \displaybreak[0]\\[2ex]
\bar{A}_{\pm}&=(E_0-S_3)(m_{\psi}\mp p_3-S_3)\,, \displaybreak[0]\\[2ex]
\bar{B}_{\pm}&=\pm(p_1+\mathrm{i}p_2)(m_{\psi}\pm p_3-S_3)\,,
\end{align}
with the normalization factor
\begin{equation}
\bar{N}_u^{(1)}=\frac{1}{4\sqrt{S_3(E_0-S_3)(S_3-m_{\psi})}}\left(2+\frac{S_3}{E_0^2}\hat{\mathcal{A}}^3\right)\,.
\end{equation}
\end{subequations}
Here, the asterisk means complex conjugation and $\varepsilon$ is the matrix representation of the two-dimensional Levi-Civita symbol.
Hence, the spinors can again be expressed completely in terms of two functions, where additional sign choices must be taken into
account. The Lorentz-violating contribution involves a global ratio of certain momentum components over the standard particle energy
squared.

\subsubsection{Second particle spinor and antiparticle spinors}
\label{sec:remaining-spinors-pseudovector}

For the pseudovector operator $\hat{\mathcal{A}}^{\mu}$, both particle spinors are related to each other by changing the sign of certain
quantities. For the isotropic case, the latter is the magnitude $|\mathbf{p}|$ of the momentum and for the anisotropic cases, these are the
quantities $S_i$:
\begin{subequations}
\begin{align}
E_u^{(-)}|^{\bigcirc}(|\mathbf{p}|)&=E_u^{(+)}|^{\bigcirc}(-|\mathbf{p}|)\,,\quad u^{(2)}|_{E_u^{(-)}}^{\bigcirc}(|\mathbf{p}|)=u^{(1)}|_{E_u^{(+)}}^{\bigcirc}(-|\mathbf{p}|)\,, \\[2ex]
E_u^{(-)}|^{\hat{\mathcal{A}}^i}(S_i)&=E_u^{(+)}|^{\hat{\mathcal{A}}^i}(-S_i)\,,\quad u^{(2)}|_{E_u^{(-)}}^{\hat{\mathcal{A}}^i}(S_i)=u^{(1)}|_{E_u^{(+)}}^{\hat{\mathcal{A}}^i}(-S_i)\,.
\end{align}
\end{subequations}
Hence, the objects $|\mathbf{p}|$ and $S_i$ are essential as they control both the particle energies and the spinor types. The
antiparticle spinors can be computed from the particle spinors by applying the charge conjugation matrix $C$ to them, cf.~Eq.~(21) of
\cite{Kostelecky:2000mm}. Independently of any representation, the charge-conjugated spinor reads $\psi_c=C\overline{\psi}^T$,
where $C=B(\gamma^0)^T$ with $-(\gamma^{\mu})^{*}=B^{-1}\gamma^{\mu}B$. In the chiral representation, we find that
$B=\mathrm{i}\gamma^2$ and thus $\psi_c=\mathrm{i}\gamma^2[(\gamma^0)^T]^2\psi^{*}=\mathrm{i}\gamma^2\psi^{*}$.
Therefore, the charge-conjugated spinors in the special frameworks considered result in
\begin{equation}
\label{eq:antiparticle-spinors-pseudovector}
v^{(1,2)}|_{E_v^{(\pm)}}=\begin{pmatrix}
u^{(2,1)}_4 \\
-u^{(2,1)}_3 \\
-u^{(2,1)}_2 \\
u^{(2,1)}_1 \\
\end{pmatrix}^{*}(-d^{\mu\nu\dots})\,,\quad N_v^{(1,2)}=N_u^{(2,1)}(-d^{\mu\nu\dots})\,,
\end{equation}
where $N_v^{(1,2)}$ are the spinor normalization constants for the antiparticle spinors. So the spinor components just have to be
rearranged including additional signs at appropriate positions (followed by a complex conjugation). Recall that the signs of the $d$
coefficients have to be reversed as well, since the latter are odd under charge conjugation, cf.~Table P31 in \cite{Kostelecky:2008ts}.

\subsection{Energies and spinors for $\boldsymbol{\hat{\mathcal{T}}^{\mu\nu}}$}

The second type of frameworks that shall be considered are based on the two-tensor operator $\hat{\mathcal{T}}^{\mu\nu}$ that comprises
both the $H$ and $g$ coefficients. The structure of the spinors is expected to be more complicated than the previous results for
$\hat{\mathcal{A}}^{\mu}$. We consider only one of the six nonzero components of $\hat{\mathcal{T}}^{\mu\nu}$ at a time.

\subsubsection{Operator $\hat{\mathcal{T}}^{0i}$ with $i=\{1,2\}$}
\label{eq:first-spinor-pseudoscalar-mixed}

The spinor results for $\hat{\mathcal{T}}^{0i}$, with $i=\{1,2\}$, are observed to be related to each other. Once the first particle
spinor for one of the two operators is known, the spinor for the other is obtained by rearranging its components and by relabelling
the momentum components. A similar behavior was observed for the anisotropic $\hat{\mathcal{A}}^{1,2}$ in \secref{sec:operator-ahat-12}.
First, the modified particle dispersion relations read
\begin{equation}
\label{eq:particle-energies-T01-T02}
E_u^{(\pm)}|^{\hat{\mathcal{T}^{0i}}}=E_0\left(1\pm\frac{S_i}{E_0^2}\hat{\mathcal{T}}^{0i}\right)\,,\quad S_i=\sqrt{\mathbf{p}^2-p_i^2}\,.
\end{equation}
In contrast to the quantities $S_i$
that were defined in the context of the pseudoscalar operator $\hat{\mathcal{A}}^{\mu}$, such as in \eqref{eq:energy-pseudovector},
the current $S_i$ do not depend on the fermion mass, but only on certain momentum components.
The spinors associated to the particle energy $E_u^{(+)}$ for both operators are based on a single master function $\hat{U}$
and they are given by:
\begin{subequations}
\begin{align}
\label{eq:particle-spinor-h01}
u^{(1)}|_{E_u^{(+)}}^{\hat{\mathcal{T}}^{01}}&=\hat{N}_u^{(1)}\hat{U}\left(\mathbf{p},\hat{\mathcal{T}}^{01},S_1\right)\,, \displaybreak[0]\\[2ex]
\label{eq:particle-spinor-h02}
u^{(1)}|^{\hat{\mathcal{T}}^{02}}_{E_u^{(+)}}&=\hat{N}^{(1)}_u\begin{pmatrix}
-\mathrm{i}\hat{U}_1 \\
\hat{U}_2 \\
-\mathrm{i}\hat{U}_3 \\
\hat{U}_4 \\
\end{pmatrix}\left(p_1=p_2,p_2=-p_1,p_3,\hat{\mathcal{T}}^{02},S_2\right)\,, \displaybreak[0]\\[2ex]
\hat{U}(\mathbf{p},X,S_i)&=\begin{pmatrix}
\boldsymbol{\hat{\phi}}_+ \\
\boldsymbol{\hat{\phi}}_- \\
\end{pmatrix}+\frac{1}{2E_0^2}\begin{pmatrix}
\delta\boldsymbol{\hat{\phi}}_+ \\
\delta\boldsymbol{\hat{\phi}}_- \\
\end{pmatrix}X\,, \displaybreak[0]\\[2ex]
\boldsymbol{\hat{\phi}}_{\pm}&=\begin{pmatrix}
\hat{A}_{\mp} \\
\hat{B}_{\pm} \\
\end{pmatrix}\,, \quad
\delta\boldsymbol{\hat{\phi}}_{\pm}=\mathrm{i}p_1\begin{pmatrix}
\hat{A}_{\pm} \\
\hat{B}_{\mp} \\
\end{pmatrix}\pm\mathrm{i}m_{\psi}\begin{pmatrix}
\hat{B}_{\pm} \\
\hat{A}_{\mp} \\
\end{pmatrix}\,, \displaybreak[0]\\[2ex]
\hat{A}_{\pm}&=\mathrm{i}(E_0+m_{\psi})(p_2-S_i)\pm p_3(p_1-\mathrm{i}S_i)\,, \displaybreak[0]\\[2ex]
\hat{B}_{\pm}&=p_3(E_0+m_{\psi}\pm p_3)\pm(p_2-\mathrm{i}p_1)(p_2-S_i)\,,
\end{align}
with the normalization factor
\begin{equation}
\hat{N}_u^{(1)}=\frac{1}{4}\sqrt{\frac{S_i+p_2}{p_3^2S_i(E_0+m_{\psi})}}\left(2+\frac{S_i}{E_0^2}\hat{\mathcal{T}}^{0i}\right)\,.
\end{equation}
\end{subequations}
Again, the transformation $p_1=p_2$, $p_2=-p_1$ must be applied to the normalization factor as well,
but the momentum components within the Lorentz-violating operator should not be touched,
cf.~\secref{sec:operator-ahat-12}. Furthermore, we have observed that these results
are very similar to the spinors for $\hat{\mathcal{A}}^{1,2}$ in structure. There are the following
correspondences between the parameters that appear in both spinors:
\begin{subequations}
\begin{align}
\hat{A}_{1,2}&=\mathrm{i}\breve{A}_{2,1}(p_1\leftrightarrow \pm p_2,p_3\leftrightarrow \pm m_{\psi})\,, \\[2ex]
\hat{B}_{1,2}&=\pm\breve{A}_{2,1}(p_1\mapsto \pm p_2,p_2\mapsto \mp p_1,p_3\mapsto \mp m_{\psi},m_{\psi}\mapsto \pm p_3)\,.
\end{align}
\end{subequations}

\subsubsection{Operator $\hat{\mathcal{T}}^{03}$}

The particle energies for the operator $\hat{\mathcal{T}}^{03}$ are given by \eqref{eq:particle-energies-T01-T02}
with $i=3$. The first particle spinor is simpler than the previous two:
\begin{subequations}
\begin{align}
\label{eq:particle-spinor-h03}
u^{(1)}|_{E_u^{(+)}}^{\hat{\mathcal{T}}^{03}}&=\tilde{N}^{(1)}_u\tilde{U}\left(\mathbf{p},\hat{\mathcal{T}}^{03},S_3\right)\,, \displaybreak[0]\\[2ex]
\tilde{U}(\mathbf{p},X,S_3)&=\begin{pmatrix}
\boldsymbol{\tilde{\phi}}_+ \\
\boldsymbol{\tilde{\phi}}_- \\
\end{pmatrix}+\frac{1}{2E_0^2}\begin{pmatrix}
\delta\boldsymbol{\tilde{\phi}}_- \\
\delta\boldsymbol{\tilde{\phi}}_+ \\
\end{pmatrix}X\,, \displaybreak[0]\\[2ex]
\boldsymbol{\tilde{\phi}}_{\pm}&=\begin{pmatrix}
(\mathrm{i}p_1+p_2)\tilde{A}_{\pm} \\
S_3\tilde{A}_{\mp} \\
\end{pmatrix}\,, \quad
\delta\boldsymbol{\tilde{\phi}}_{\pm}=\begin{pmatrix}
(\mathrm{i}p_1+p_2)\tilde{B}_{\pm} \\
S_3\tilde{B}_{\mp} \\
\end{pmatrix}\,, \displaybreak[0]\\[2ex]
\tilde{A}_{\pm}&=E_0\pm \mathrm{i}S_3\mp p_3+m_{\psi}\,, \displaybreak[0]\\[2ex]
\tilde{B}_{\pm}&=\mathrm{i}p_3(\tilde{A}_{\pm}-m_{\psi})+\mathrm{i}m_{\psi}(p_3\mp\tilde{A}_{\mp})\,,
\end{align}
with the normalization constant
\begin{equation}
\tilde{N}_u^{(1)}=\frac{1}{4|S_3|\sqrt{E_0+m_{\psi}}}\left(2+\frac{S_3}{E_0^2}\hat{\mathcal{T}}^{03}\right)\,.
\end{equation}
\end{subequations}
Two functions including sign choices are sufficient to parameterize the solution of the Dirac equation.
Note the absolute-value bars around $S_3>0$ in the normalization. They are stated explicitly to indicate
that the normalization does not change sign for the antiparticle spinors when $S_3$ is replaced by
$-S_3$, cf.~\secref{sec:second-particle-spinor-tmunu} below. This is also the only case where a
quantity $S_i$ appears outside of a square root function in the standard part of the normalization factor.

\subsubsection{Operator $\hat{\mathcal{T}}^{i3}$ with $i=\{1,2\}$}

Last but not least, we want to state the spinor solutions for the $\hat{\mathcal{T}}^{\mu\nu}$ operator with
two spatial indices nonvanishing. The subsequent results are simpler than the previous ones for $\hat{\mathcal{T}}^{0i}$
with one timelike index. First of all, the energy levels at first order in Lorentz violation can be expressed as
before, with anisotropic quantities $S_{ij}$ that here depend on the particle mass:
\begin{equation}
\label{eq:energies-pseudoscalar-spatial-1}
E_u^{(\pm)}|^{\hat{\mathcal{T}}^{ij}}=E_0\left(1\pm\frac{S_{ij}}{E_0^2}\hat{\mathcal{T}}^{ij}\right)\,,\quad S_{ij}=\sqrt{p_i^2+p_j^2+m_{\psi}^2}\,.
\end{equation}
The spinors for $\hat{\mathcal{T}}^{13}$ and $\hat{\mathcal{T}}^{23}$ are again related to each other,
which makes it possible to express them via a single master function $\bar{\bar{U}}$. A similar behavior
was encountered for $\hat{\mathcal{A}}^{1,2}$ in \secref{sec:operator-ahat-12}, and for
$\hat{\mathcal{T}}^{01,02}$ in cf.~\secref{eq:first-spinor-pseudoscalar-mixed}.
\begin{subequations}
\begin{align}
\label{eq:particle-spinor-h23}
u^{(1)}|^{\hat{\mathcal{T}}^{23}}_{E_u^{(+)}}&=\bar{\bar{N}}_u^{(1)}\bar{\bar{U}}(\mathbf{p},\hat{\mathcal{T}}^{23},S_{23})\,, \displaybreak[0]\\[2ex]
\label{eq:particle-spinor-h13}
u^{(1)}|^{\hat{\mathcal{T}}^{13}}_{E_u^{(+)}}&=\bar{\bar{N}}^{(1)}_u\begin{pmatrix}
\mathrm{i}\bar{\bar{U}}_1 \\
\bar{\bar{U}}_2 \\
\mathrm{i}\bar{\bar{U}}_3 \\
\bar{\bar{U}}_4 \\
\end{pmatrix}\left(p_1=-p_2,p_2=p_1,p_3,\hat{\mathcal{T}}^{13},S_{13}\right)\,, \displaybreak[0]\\[2ex]
\bar{\bar{U}}(\mathbf{p},X,S_{ij})&=\begin{pmatrix}
\boldsymbol{\bar{\bar{\phi}}}_+ \\
\boldsymbol{\bar{\bar{\phi}}}_- \\
\end{pmatrix}+\frac{p_1}{2E_0^2}\begin{pmatrix}
-\boldsymbol{\bar{\bar{\phi}}}_- \\
\boldsymbol{\bar{\bar{\phi}}}_+ \\
\end{pmatrix}X\,, \displaybreak[0]\\[2ex]
\boldsymbol{\bar{\bar{\phi}}}_{\pm}&=\pm\begin{pmatrix}
-\bar{\bar{A}}_{\pm} \\
\bar{\bar{B}}_{\pm} \\
\end{pmatrix}\,, \displaybreak[0]\\[2ex]
\bar{\bar{A}}&=\frac{p_1p_3+(E_0+m_{\psi})S_{ij}}{E_0(E_0+m_{\psi})-p_1(p_1+\mathrm{i}p_2)}\,, \displaybreak[0]\\[2ex]
\bar{\bar{A}}_{\pm}&=p_1-\mathrm{i}p_2\pm(E_0+m_{\psi}\mp p_3)\bar{\bar{A}}\,, \displaybreak[0]\\[2ex]
\bar{\bar{B}}_{\pm}&=(p_1+\mathrm{i}p_2)\bar{\bar{A}}\pm(E_0+m_{\psi}\pm p_3)\,,
\end{align}
with the normalization factor
\begin{equation}
\bar{\bar{N}}_u^{(1)}=\frac{\sqrt{E_0+m_{\psi}-p_1p_3/S_{ij}}}{4(E_0+m_{\psi})}\left(2+\frac{S_{ij}}{E_0^2}\hat{\mathcal{T}}^{ij}\right)\,.
\end{equation}
\end{subequations}
In contrast to the spinors for $\hat{\mathcal{T}}^{0i}$, only two two-dimensional vectors are sufficient to construct
the spinors at first order in Lorentz violation. The same will be true for $\hat{\mathcal{T}}^{12}$ below.

\subsubsection{Operator $\hat{\mathcal{T}}^{12}$}

For the last operator to be considered, the energies are given by \eqref{eq:energies-pseudoscalar-spatial-1} with
$\{i,j\}=\{1,2\}$. The first particle spinor reads
\begin{subequations}
\begin{align}
\label{eq:particle-spinor-h12}
u^{(1)}|^{\hat{\mathcal{T}}^{12}}_{E_u^{(+)}}&=\check{N}_u^{(1)}\check{U}(\mathbf{p},\hat{\mathcal{T}}^{12},S_{12})\,, \displaybreak[0]\\[2ex]
\check{U}(\mathbf{p},X,S_i)&=\begin{pmatrix}
\boldsymbol{\check{\phi}}_+ \\
\boldsymbol{\check{\phi}}_- \\
\end{pmatrix}+\frac{p_3}{2E_0^2}\begin{pmatrix}
-\boldsymbol{\check{\phi}}_- \\
\boldsymbol{\check{\phi}}_+ \\
\end{pmatrix}X\,, \displaybreak[0]\\[2ex]
\boldsymbol{\check{\phi}}_{\pm}&=\pm\begin{pmatrix}
-\check{A}_{\pm} \\
\check{B}_{\pm} \\
\end{pmatrix}\,, \displaybreak[0]\\[2ex]
\check{A}&=\frac{(E_0+m_{\psi})(E_0-S_{12})-p_3^2}{(p_1+\mathrm{i}p_2)p_3}\,, \displaybreak[0]\\[2ex]
\check{A}_{\pm}&=p_1-\mathrm{i}p_2\pm (E_0+m_{\psi}\mp p_3)\check{A}\,, \displaybreak[0]\\[2ex]
\check{B}_{\pm}&=(p_1+\mathrm{i}p_2)\check{A}\pm (E_0+m_{\psi}\pm p_3)\,, \displaybreak[0]\\[2ex]
\check{N}_u^{(1)}&=\frac{\sqrt{E_0+m_{\psi}+S_{12}+E_0m_{\psi}/S_{12}}}{4(E_0+m_{\psi})}\left(2+\frac{S_{12}}{E_0^2}\hat{\mathcal{T}}^{12}\right)\,.
\end{align}
\end{subequations}
Evidently, the form of the spinor is analogous to the results for $\hat{\mathcal{T}}^{01}$ and
$\hat{\mathcal{T}}^{02}$. Furthermore, $\check{A}_{\pm}$, $\check{B}_{\pm}$ have the same form as the
previous $\bar{\bar{A}}_{\pm}$, $\bar{\bar{A}}_{\pm}$. However, $\check{A}\neq \bar{\bar{A}}$ and the
structure of the normalization factors differs from each other, too.

\subsubsection{Second particle spinor and antiparticle spinors}
\label{sec:second-particle-spinor-tmunu}

\begin{table}[t]
\setlength\extrarowheight{5pt}
\begin{tabular}{cccc}
\toprule
Coeff. $X$ & $E_u^{(\pm)}/E_0$ & Spinors & Definitions \\
\colrule
$(4/3)E_0\hat{\ring{d}}-\hat{\ring{b}}+m_{\psi}\hat{\ring{g}}$ & $1\pm (|\mathbf{p}|/E_0^2)X$ & \eqref{eq:particle-spinors-b0} & \\
\colrule
$\hat{\mathcal{A}}^1$ & $1\pm (S_i/E_0^2)X$ & \eqref{eq:particle-spinors-b1} & $S_i\equiv\sqrt{p_i^2+m_{\psi}^2}$ \\
$\hat{\mathcal{A}}^2$ &                     & \eqref{eq:particle-spinors-b2} & \\
$\hat{\mathcal{A}}^3$ &                     & \eqref{eq:particle-spinors-b3} & \\
\colrule
$\hat{\mathcal{T}}^{01}$ & $1\pm (S_i/E_0^2)X$ & \eqref{eq:particle-spinor-h01} & $S_i\equiv\sqrt{\mathbf{p}^2-p_i^2}$ \\
$\hat{\mathcal{T}}^{02}$ &                          & \eqref{eq:particle-spinor-h02} & \\
$\hat{\mathcal{T}}^{03}$ & & \eqref{eq:particle-spinor-h03} & \\
\colrule
$\hat{\mathcal{T}}^{13}$ & $1\pm(S_{ij}/E_0^2)X$ & \eqref{eq:particle-spinor-h13} & $S_{ij}\equiv\sqrt{p_i^2+p_j^2+m_{\psi}^2}$ \\
$\hat{\mathcal{T}}^{23}$ &                         & \eqref{eq:particle-spinor-h23} & \\
$\hat{\mathcal{T}}^{12}$ & & \eqref{eq:particle-spinor-h12} & \\
\botrule
\end{tabular}
\caption{Collection of modified particle energies and spinors.}
\label{tab:summary-results}
\end{table}
For the tensor operator $\hat{\mathcal{T}}^{\mu\nu}$ the first particle spinors only were given previously. The second particle spinor
can be obtained from the first in an easy way in replacing all quantities $S_i$ or $S_{ij}$ by their counterparts with opposite sign:
\begin{subequations}
\begin{align}
E_u^{(-)}|^{\hat{\mathcal{T}}^{0i}}(S_i)&=E_u^{(+)}|^{\hat{\mathcal{T}}^{0i}}(-S_i)\,,\quad E_u^{(-)}|^{\hat{\mathcal{T}}^{ij}}(S_{ij})=E_u^{(+)}|^{\hat{\mathcal{T}}^{ij}}(-S_{ij})\,, \\[2ex]
u^{(2)}|_{E_u^{(-)}}^{\hat{\mathcal{T}}^{0i}}(S_i)&=u^{(1)}|_{E_u^{(+)}}^{\hat{\mathcal{T}}^{0i}}(-S_i)\,,\quad u^{(2)}|_{E_u^{(-)}}^{\hat{\mathcal{T}}^{ij}}(S_{ij})=u^{(1)}|_{E_u^{(+)}}^{\hat{\mathcal{T}}^{ij}}(-S_{ij})\,.
\end{align}
\end{subequations}
In analogy to the pseudovector operator, the types of energy and spinor are controlled by the quantities $S_i$ and $S_{ij}$. Also, the antiparticle
spinors for $\hat{\mathcal{T}}^{\mu\nu}$ are determined completely by the components of the particle spinors. With the charge-conjugated
spinor $\psi_c=\mathrm{i}\gamma^2\psi^{*}$ in the chiral representation, we obtain them directly from Eq.~(21) of
\cite{Kostelecky:2000mm}:
\begin{equation}
\label{eq:antiparticle-spinors-two-tensor}
v^{(1,2)}|_{E_v^{(\pm)}}=\begin{pmatrix}
u^{(2,1)}_4 \\
-u^{(2,1)}_3 \\
-u^{(2,1)}_2 \\
u^{(2,1)}_1 \\
\end{pmatrix}^{*}(-H^{\mu\nu\dots})\,,\quad N_v^{(1,2)}=N_u^{(2,1)}(-H^{\mu\nu\dots})\,.
\end{equation}
The signs of the $H$ coefficients also have to be reversed, since these are odd under charge conjugation, cf.~Table P31 in
\cite{Kostelecky:2008ts}.

\subsection{Additional observations}

It is interesting to note that some of the spinors can be cast into a very simple shape by factoring the Lorentz-invariant
contribution out of each component. This can be carried out for the ``isotropic case,'' for each of the spacelike components
of the pseudovector operator $\hat{\mathcal{A}}^{\mu}$, and for the two-tensor operator components $\hat{\mathcal{T}}^{03}$,
$\hat{\mathcal{T}}^{12}$. For these cases, it is possible to express each spinor component of the first particle spinor in
the following form:
\begin{subequations}
\begin{align}
u^{(1)}_i|_{E_u^{(+)}}^{\bigcirc}&=\begin{pmatrix}
\boldsymbol{\ring{\phi}}_{-} \\
\boldsymbol{\ring{\chi}}_{+} \\
\end{pmatrix}_i(1-\theta\mathscr{V}_i)\,,\quad \theta=\frac{1}{2E_0^2}\left(\hat{\ring{\mathcal{A}}}-\hat{\ring{g}}\frac{\mathbf{p}^2}{m_{\psi}}\right)\,, \displaybreak[0]\\[2ex]
u^{(1)}_i|_{E_u^{(+)}}^{\hat{\mathcal{A}}^1}&=\begin{pmatrix}
\boldsymbol{\breve{\phi}}_+ \\
\boldsymbol{\breve{\phi}}_- \\
\end{pmatrix}_i(1-\iota\mathscr{W}_i)\,,\quad \iota=\frac{\hat{\mathcal{A}}^1}{2E_0}\left(\frac{S_1}{E_0}p_3+\mathrm{i}p_2\right)\,, \displaybreak[0]\\[2ex]
u^{(1)}_i|_{E_u^{(+)}}^{\hat{\mathcal{A}}^3}&=\begin{pmatrix}
\boldsymbol{\bar{\phi}}_+ \\
\boldsymbol{\bar{\phi}}_- \\
\end{pmatrix}_i(1-\kappa\mathscr{X}_i)\,,\quad \kappa=\frac{\hat{\mathcal{A}}^3}{2E_0^2}\,, \displaybreak[0]\\[2ex]
u^{(1)}_i|_{E_u^{(+)}}^{\hat{\mathcal{T}}^{03}}&=\begin{pmatrix}
\boldsymbol{\tilde{\phi}}_+ \\
\boldsymbol{\tilde{\phi}}_- \\
\end{pmatrix}_i(1-\tau\mathscr{W}_i)\,,\quad \tau=\frac{\hat{\mathcal{T}}^{03}}{2E_0}\left(\frac{S_3}{E_0}p_3+\mathrm{i}m_{\psi}\right)\,, \displaybreak[0]\\[2ex]
u^{(1)}_i|_{E_u^{(+)}}^{\hat{\mathcal{T}}^{12}}&=\begin{pmatrix}
\boldsymbol{\check{\phi}}_+ \\
\boldsymbol{\check{\phi}}_- \\
\end{pmatrix}_i(1-\omega\mathscr{W}_i)\,,\quad \omega=\hat{\mathcal{T}}^{12}\frac{S}{2E_0^2}p_3\,,
\end{align}
with the vectors
\begin{align}
\mathscr{V}&=\begin{pmatrix}
|\mathbf{p}|+E_0 \\
|\mathbf{p}|+E_0 \\
|\mathbf{p}|-E_0 \\
|\mathbf{p}|-E_0 \\
\end{pmatrix}\,,\quad \mathscr{W}=\begin{pmatrix}
1/(p_3-E_0) \\
1/(p_3+E_0) \\
1/(p_3+E_0) \\
1/(p_3-E_0) \\
\end{pmatrix}\,,\quad \mathscr{X}=\begin{pmatrix}
S+E_0 \\
S-E_0 \\
S+E_0 \\
S-E_0 \\
\end{pmatrix}\,.
\end{align}
\end{subequations}
Note the indices $i$ on both sides of the relations, which refers to a particular component $i$ of the spinors and defined vectors,
respectively. What is characteristic for the stated spinors is the exceptionally simple form of the Lorentz-violating terms. For
the second and the fourth of these spinors only, the Lorentz-violating contribution has an imaginary part. The spinor for
$\hat{\mathcal{A}}^2$ can be obtained from the spinor for $\hat{\mathcal{A}}^1$, cf.~\secref{sec:operator-ahat-12}, whereby it is
discarded here. For the remaining spinors, it was not possible to obtain results with an analog simplicity, which is why they
are dropped as well. Besides, the ratios,
\begin{equation}
\Omega_u\equiv\sqrt{\frac{u^{\dagger}u|^{\hat{X}^{\mu\dots}=0}}{u^{\dagger}u}}\,,\quad \Omega_v\equiv\sqrt{\frac{v^{\dagger}v|^{\hat{X}^{\mu\dots}=0}}{v^{\dagger}v}}\,,
\end{equation}
with a generic Lorentz-violating operator $\hat{X}^{\mu\dots}$, are further interesting quantities. They are measures for how much
the Lorentz-violating part of the spinors is suppressed compared to the Lorentz-invariant part. The quantities $\Omega_u$ are simple
for all operators considered and they read
\begin{subequations}
\begin{align}
\Omega_u|^{\bigcirc}&=\frac{m_{\psi}}{2E_0^2}\left|\hat{\ring{\mathcal{A}}}-\hat{\ring{g}}\frac{\mathbf{p}^2}{m_{\psi}}\right|\,,\quad \Omega_u|^{\hat{\mathcal{A}}^i}=\frac{|\hat{\mathcal{A}}^i|}{2E_0^2}\sqrt{\mathbf{p}^2-p_i^2}\,, \displaybreak[0]\\[2ex]
\Omega_u|^{\hat{\mathcal{T}}^{0i}}&=\frac{|\hat{\mathcal{T}}^{0i}|}{2E_0^2}\sqrt{p_i^2+m_{\psi}^2}\,,\quad \Omega_u|^{\hat{\mathcal{T}}^{ij}}=\frac{|\hat{\mathcal{T}}^{ij}\varepsilon^{ijk}p_k|}{2E_0^2}\,,
\end{align}
\end{subequations}
where in the latter equal indices are not summed over. The results for both particle spinors correspond to each other.
Furthermore, $\Omega_v=\Omega_u(-d^{\mu\nu},-H^{\mu\nu})$ for antiparticles. The findings tell us that Lorentz-violating effects may be additionally
suppressed for $|\mathbf{p}|\ll m_{\psi}$, i.e., for decreasing momentum. First of all, all Lorentz-violating operators have mass
dimension 1. As $\Omega_u$ is dimensionless, the operator has to be divided by a dimensionful scale, where for vanishing momentum the
only one is the fermion mass. This holds for the ``isotropic case'' and the two-tensor operator with one timelike index.
For the remaining frameworks, there is an additional suppression factor $p_i/m_{\psi}$ with the largest momentum component $p_i$.

Finally, another issue related to the spinors shall be mentioned at this point. There is a matrix transformation that connects the chiral representation
of $\gamma$ matrices to the Dirac representation~\cite{Nagashima:2010}. It can be written in the form $M_D=VM_{\mathrm{ch}}V^{-1}$,
where $M_{\mathrm{ch}}$ and $M_D$ are $\gamma$ matrices in the chiral and the Dirac representation, respectively. The matrix $V$
here corresponds to the matrix $V$ in \eqref{eq:transformation-matrices}, contained in the transformation that diagonalizes the Dirac
operator (with $\gamma^0$ and $\gamma^5$ themselves in the chiral representation). The spinors of both representations are then
related by $\psi_D=V\psi_{\mathrm{ch}}$. Hence, all the spinors obtained in this paper can be transformed from the chiral to the
Dirac representation according to this rule. We carried this out and considered the limits of the spinors for zero Lorentz
violation, where the particle and antiparticle states become spin-degenerate. It was verified successfully that there exist proper
linear combinations of the particle spinors $u^{(\alpha)}$ and of the antiparticle spinors $v^{(\alpha)}$, such that the standard
solutions of the Dirac equation are obtained:
\begin{subequations}
\begin{align}
\sum_{\alpha=1,2} \varsigma_s^{(\alpha)}u^{(\alpha)}&=\begin{pmatrix}
\boldsymbol{\phi}^{(s)} \\
\boldsymbol{\sigma}\cdot\mathbf{p}/(E_0+m_{\psi}) \cdot \boldsymbol{\phi}^{(s)} \\
\end{pmatrix}\,,\quad \boldsymbol{\phi}^{(1)}=\begin{pmatrix}
1 \\
0 \\
\end{pmatrix}\,,\quad \boldsymbol{\phi}^{(2)}=\begin{pmatrix}
0 \\
1 \\
\end{pmatrix}\,, \\[2ex]
\sum_{\alpha=1,2} \tau_s^{(\alpha)}v^{(\alpha)}&=\begin{pmatrix}
\boldsymbol{\sigma}\cdot\mathbf{p}/(E_0+m_{\psi})\cdot \boldsymbol{\chi}^{(s)} \\
\boldsymbol{\chi}^{(s)} \\
\end{pmatrix}\,,\quad \boldsymbol{\chi}^{(1)}=\begin{pmatrix}
1 \\
0 \\
\end{pmatrix}\,,\quad \boldsymbol{\chi}^{(2)}=\begin{pmatrix}
0 \\
1 \\
\end{pmatrix}\,,
\end{align}
\end{subequations}
with $\boldsymbol{\sigma}=(\sigma^1,\sigma^2,\sigma^3)$, where $\sigma^i$ are the Pauli matrices. The parameters $\varsigma_s^{(\alpha)}$
and $\tau_s^{(\alpha)}$ are chosen appropriately.

\section{Spinor matrices and optical theorem}
\label{sec:spinor-matrices-optical-theorem}
\setcounter{equation}{0}

For calculations in high-energy physics that are based on quantum field theory, the spinors are often not needed directly.
Instead, when computing matrix element squares spinors are always combined in the form of $u\overline{u}$, where
the latter is a $4\times 4$ matrix in spinor space. In the Lorentz-invariant case, it often suffices to calculate unpolarized
cross sections, which means that one has to average over initial particle spins. As such a scattering or decay process is a
quantum process, the final state is not predictable, which is why the final particle spins must be summed over and the phase
space has to be integrated out. Hence, in the standard case the following well-known expressions (with proper normalization
of the spinors) are usually extremely helpful:
\begin{subequations}
\label{eq:spinor-matrices-standard}
\begin{align}
\sum_{\alpha=1,2} u^{(\alpha)}\overline{u}^{(\alpha)}&=\cancel{p}+m_{\psi}\mathds{1}_4\,, \\[2ex]
\sum_{\alpha=1,2} v^{(\alpha)}\overline{v}^{(\alpha)}&=\cancel{p}-m_{\psi}\mathds{1}_4\,.
\end{align}
\end{subequations}
For the Lorentz-violating case with broken spin degeneracy, there are essential differences. First, due to the modified
kinematics in Lorentz-violating frameworks, a small part of the phase space of otherwise forbidden particle processes may open
and render them possible. Examples are Cherenkov-type processes in vacuo and decays of
photons into electron-positron pairs. Since in a theory with broken spin degeneracy both the particles and antiparticles
can have two distinct energies, such a process may be allowed only for one of these energies. Under this circumstance, sums
over both particle spins in the matrix element square do not have to be carried out as only a single spin state contributes. So what we are
interested in are expressions such as $u^{(\alpha)}\overline{u}^{(\alpha)}$ or $v^{(\alpha)}\overline{v}^{(\alpha)}$,
with a particular $\alpha$. From now on these matrices will be referred to as ``spinor matrices.''
However, for calculational purposes, the matrices themselves are not that useful. Instead, such a matrix should be
expressed in terms of the 16 Dirac bilinears. The matrix structure is supposedly more complicated than what we encountered
in \eqref{eq:spinor-matrices-standard}, where the Dirac matrices and the unit matrix only appear. For the Lorentz-violating
frameworks considered here, the matrix structure is expected to include a large number of the 16 Dirac bilinears such as for
the propagators, cf.~Eqs.~(\ref{eq:propagator-pseudovector}), (\ref{eq:propagator-two-tensor}).

There are two possibilities of computing the spinor matrices. The first takes into account the spinors that were determined
in the previous section. The matrices are calculated directly based on these spinors and they are expressed in terms
of the Dirac bilinears. The resulting expressions will be valid at first order in Lorentz violation. The second possibility
is to use the optical theorem. The latter gives a relationship between the imaginary part of a forward
scattering amplitude to the total cross sections of all processes that are energetically allowed by cutting the propagators
in the diagram representing the forward scattering amplitude.
The validity of the optical theorem was demonstrated at tree-level for various sets of Lorentz-violating frameworks,
such as both minimal and nonminimal modifications of the photon and the fermion sector
\cite{Klinkhamer:2010zs,Schreck:2011ai,Schreck:2013gma,Schreck:2013kja,Maniatis:2014xja,Schreck:2014qka}. Therefore, it
is not expected to lose its validity for the range of coefficients considered within the current paper. Nevertheless, a
cross check will be carried out at first order in Lorentz violation using the spinor matrices obtained from the spinors
directly.
\begin{figure}
\centering
\includegraphics{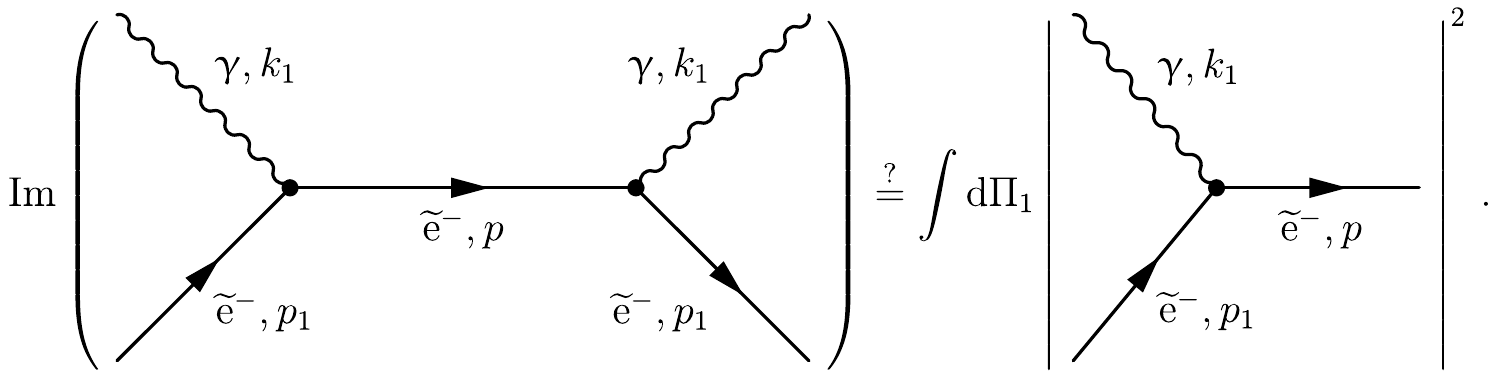}
\caption{Forward scattering amplitude of one of the two contributions for Compton scattering that is linked to the
total cross section of photon absorption by an electron.}
\label{fig:optical-theorem}
\end{figure}

To make use of the optical theorem, a particular scattering process is needed. Thus, we have to introduce an interaction
and we decide to couple the modified fermions to standard photons. In this context it is reasonable to mention the very
recent paper \cite{Ding:2016lwt}, where nonminimal
Lorentz violation in the fermion sector is constrained by Penning trap experiments. To do so, fermions are minimally
coupled to electromagnetic fields. This amounts to replacing the particle derivative at all occurrences by the covariant
derivative $D_{\mu}=\partial_{\mu}-eA_{\mu}$, where $A_{\mu}$ is the vector potential and $e$ the elementary charge. However,
in contrast to the partial derivative, the covariant derivative does not commute, which introduces additional interactions
for the nonminimal terms. In principle, when computing the cross section or decay rate of a particular process, all these
additional interactions must be taken into account. Since we are only interested in investigating the validity and implications
of the optical theorem for fermions and since this calculation will be purely formal, the exact form of the interaction is not
important and we can even resort to the standard one. The fermion propagator and the external particles, which are described
by spinors, lie in the center of our considerations.

The particular process to be analyzed is one of the two possible contributions of Compton scattering with an incoming photon
and electron scattering and producing another photon-electron pair, cf.~\figref{fig:optical-theorem}. The second contribution with both vertices interchanged is not needed in this formal
analysis. The left-hand side of the equation, presented in the latter figure, contains the forward scattering amplitude
of this process, where the initial and the final state correspond to each other. Based on the optical theorem, the imaginary
part of the left-hand side is linked to the total cross section of a photon absorption process shown on the right-hand
side. The left-hand side of the equation contains a propagator where the right-hand side involves external fermions and,
therefore, spinors that are combined into spinor matrices.

We consider the incoming electron to be in the spin-state associated to the energy $E_u^{(+)}$; it is then described by the
spinor $u^{(1)}$. The standard Feynman rules for the external photons and the vertices can be employed to write up the
matrix element $\mathcal{M}\equiv \mathcal{M}(\mathrm{e^-}\upgamma\rightarrow\mathrm{e^-}\upgamma)$ of the forward
scattering amplitude (cf.~also \cite{Peskin:1995}):
\begin{equation}
\label{eq:forward-scattering-amplitude}
\mathcal{M}=-\int \frac{\mathrm{d}^4p}{(2\pi)^4}\,\delta^{(4)}(k_1+p_1-p)e^2\overline{u}^{(\alpha)}(p_1)\gamma^{\nu}S(p)\gamma^{\mu}u^{(\alpha)}(p_1)\varepsilon^{(\lambda)}_{\mu}(k_1)\varepsilon^{*(\lambda)}_{\nu}(k_1)\,,
\end{equation}
where Lorentz violation just sits in the modified fermion propagator $S$. Furthermore,
$\varepsilon^{(\lambda)}$ is the standard polarization vector of a photon in the polarization state $\lambda$.
The $\delta$ function directly behind the integration measure enforces energy-momentum conservation.
The fermion propagator is either taken from \eqref{eq:propagator-pseudovector} or \eqref{eq:propagator-two-tensor}, dependent
on whether Lorentz violation based on $\hat{\mathcal{A}}^{\mu}$ or $\hat{\mathcal{T}}^{\mu\nu}$ is considered. However,
for this formal calculation, it will be chosen generically. The Feynman-type propagator has to be used in the amplitude,
which means that an infinitesimal imaginary part must be added to the denominator:
\begin{equation}
\label{eq:propagator-poles}
\frac{1}{\Delta+\mathrm{i}\epsilon}=\frac{1}{\mathscr{Z}\left(p^0-E_u^{(+)}+\mathrm{i}\epsilon\right)\left(p^0-E_u^{(-)}+\mathrm{i}\epsilon\right)\left(p^0-E^{(+)}_<-\mathrm{i}\epsilon\right)\left(p^0-E^{(-)}_<-\mathrm{i}\epsilon\right)}\,,
\end{equation}
with a global prefactor $\mathscr{Z}$ that does not depend on $p^0$.
The imaginary number $\mathrm{i}\epsilon$ prevents singularities emerging for a vanishing $\Delta$, which is exactly what
produces the imaginary part of the forward scattering amplitude. The denominator has four roots in $p^0$, which are composed
of two positive ones $E_u^{(\pm)}$ and two negative ones $E^{(\pm)}_{<}$. The imaginary parts are added to the poles such
that the two positive poles are shifted to the lower complex half plane and the two negative poles are shifted to the upper
one. 
In \secref{sec:additional-time-derivatives}, it will be investigated under which circumstances the form of \eqref{eq:propagator-poles}
fails to be valid and how such a case can be treated. Energy-momentum conservation of the process shall
allow $\Delta$ to vanish for $p^0=E_u^{(+)}$ only. To evaluate the expression, the following identity is used
\begin{equation}
\frac{1}{p^0-E_u^{(+)}+\mathrm{i}\epsilon}=\mathcal{P}\frac{1}{p^0-E_u^{(+)}}-\mathrm{i}\pi\delta\left(p^0-E_u^{(+)}\right)\,,
\end{equation}
where $\mathcal{P}$ is the principle value. The first term is real, i.e., it does not contribute to the imaginary part
of $\mathcal{M}$. The second term does so only and it forces $p^0$ to be equal to the energy $E_u^{(+)}$ when evaluating
the integral over $p^0$. That corresponds to cutting the propagator in the Feynman diagram on the left-hand side of
\figref{fig:optical-theorem} into two lines, which represent on-shell fermions
described by the spinor $u^{(1)}$. The imaginary part of the forward scattering amplitude can then be cast into the form
\begin{subequations}
\begin{align}
\label{eq:forward-scattering-amplitude}
2\mathrm{Im}(\mathcal{M})&=\int \frac{\mathrm{d}^3p}{(2\pi)^32E_u^{(+)}}\delta^{(4)}(k_1+p_1-p)e^2\overline{u}^{(\alpha)}(p_1)\gamma^{\nu}\mathscr{C}(E_u^{(-)},E_<^{(+)},E_<^{(-)})|_{p^0=E_u^{(+)}} \notag \displaybreak[0]\\
&\phantom{{}={}\int \frac{\mathrm{d}^3p}{(2\pi)^32E_u^{(+)}}}\times\left(\widehat{\xi}^{\,\mu}\gamma_{\mu}+\widehat{\Xi}+\widehat{\Upsilon}+\widehat{\zeta}^{\,\mu}\gamma_5\gamma_{\mu}+\widehat{\psi}^{\,\mu\nu}\sigma_{\mu\nu}\right)\Big|_{p^0=E_u^{(+)}} \notag \displaybreak[0]\\
&\phantom{{}={}\int \frac{\mathrm{d}^3p}{(2\pi)^32E_u^{(+)}}}\times\gamma^{\mu}u^{(\alpha)}(p_1)\varepsilon_{\mu}^{(\lambda)}(k_1)\overline{\varepsilon}_{\nu}^{(\lambda)}(k_1)\,,
\end{align}
with the generic function
\begin{equation}
\label{eq:scalar-function}
\mathscr{C}(a,b,c)\equiv\frac{2p^0}{\mathscr{Z}(p^0-a)(p^0-b)(p^0-c)}\,.
\end{equation}
\end{subequations}
The latter function, evaluated at $p^0=E_u^{(+)}$, is the remainder of the denominator $\Delta$. The additional factor of
$2p^0$ is introduced to cancel the factor of twice the particle energy that has been generated in the denominator of the
integral measure. Thus, what remains from the propagator is the scalar function $\mathscr{C}$ and the matrix structure,
which is expressed in terms of the Dirac bilinears and which is evaluated at $p^0=E_u^{(+)}$. In principle, it is not
difficult to take into account an additional prefactor in the denominator of \eqref{eq:propagator-poles}, which is
independent of $p^0$. Such a prefactor would now appear in the denominator of \eqref{eq:scalar-function}.

Due to the validity of the optical theorem, \eqref{eq:forward-scattering-amplitude} must be equal to the total cross section
$\sigma\equiv \sigma(\mathrm{e^-}\upgamma\rightarrow \mathrm{e^-})$ of the process on the right-hand side of
\figref{fig:optical-theorem}. Thereby,
the initial electron is in the spin-state associated with the energy $E_u^{(+)}$, for which reason it is described by the
spinor $u^{(1)}$. With $\widehat{\mathcal{M}}\equiv \mathcal{M}(\mathrm{e^-}\upgamma\rightarrow \mathrm{e^-})$, we
obtain:
\begin{align}
\sigma&=\int \frac{\mathrm{d}^3p}{(2\pi)^32E_u^{(+)}}\delta^{(4)}(k_1+p_2-p) |\widehat{\mathcal{M}}|^2 \notag \displaybreak[0]\\
&=\int \frac{\mathrm{d}^3p}{(2\pi)^32E_u^{(+)}}\delta^{(4)}(k_1+p_2-p) \notag \displaybreak[0]\\
&\phantom{{}={}\int \frac{\mathrm{d}^3p}{(2\pi)^32E_u^{(+)}}}\times \left(\mathrm{i}e\overline{u}^{(1)}(p)\gamma^{\nu}u^{(\alpha)}(p_1)\varepsilon_{\nu}^{(\lambda)}(k_1)\right)^{\dagger}\mathrm{i}e\overline{u}^{(1)}(p)\gamma^{\mu}u^{(\alpha)}(p_1)\varepsilon^{(\lambda)}_{\mu}(k_1) \notag \displaybreak[0]\\
&=\int \frac{\mathrm{d}^3p}{(2\pi)^32E_u^{(+)}}\delta^{(4)}(k_1+p_1-p) \notag \displaybreak[0]\\
&\phantom{{}={}\int \frac{\mathrm{d}^3p}{(2\pi)^32E_u^{(+)}}}\times e^2\overline{u}^{(\alpha)}(p_1)\gamma^{\nu}\Big[u^{(1)}(p)\overline{u}^{(1)}(p)\Big]\gamma^{\mu} u^{(\alpha)}(p_1)\varepsilon_{\mu}^{(\lambda)}(k_1)\overline{\varepsilon}_{\nu}^{(\lambda)}(k_1)\,.
\end{align}
A comparison of the latter final result with \eqref{eq:forward-scattering-amplitude} allows for deriving an
expression for the spinor matrix formed from $u^{(1)}$. The whole computation can be performed for an electron
in the spin state connected to $E_u^{(-)}$ in an analogous way leading to the spinor matrix obtained from $u^{(2)}$.
The results will be given below. Similarly, it is possible to derive spinor matrices for the antiparticle spinor $v^{(1)}$.
To do so, the same process is considered with only the electron replaced by a positron. Then, the forward
scattering amplitude $\overline{\mathcal{M}}\equiv\overline{\mathcal{M}}(\mathrm{e^+}\upgamma\rightarrow\mathrm{e^+}\upgamma)$
reads
\begin{equation}
\label{eq:forward-scattering-amplitude-positron}
\overline{\mathcal{M}}=\int \frac{\mathrm{d}^4p}{(2\pi)^4}\,\delta^{(4)}(k_1+p_1-p)e^2\overline{v}^{(\alpha)}(p_1)\gamma^{\mu}S(-p)\gamma^{\nu}v^{(\alpha)}(p_1)\varepsilon^{(\lambda)}_{\mu}(k_1)\overline{\varepsilon}^{(\lambda)}_{\nu}(k_1)\,.
\end{equation}
The propagator has to be evaluated at the negative of the four-momentum to take into account that the momentum of the
internal line flows in a direction opposite to the arrow on this line. Besides, a global prefactor of $-1$ has to be
considered to account for the interchange of fermionic operators when applying Wick's theorem \cite{Schreck:2014qka}.
First of all, we evaluate the denominator of the propagator for the four-momentum components with their signs reversed:
\begin{subequations}
\begin{align}
\Delta_{\hat{\mathcal{A}},\hat{\mathcal{H}}}(-p)&=\left(-p^0-E_u^{(+)}\right)\left(-p^0-E_u^{(-)}\right)\left(-p^0-E_{<}^{(+)}\right)\left(-p^0-E_{<}^{(-)}\right)\big|_{\mathbf{p}\mapsto-\mathbf{p}} \notag \displaybreak[0]\\
&=\left(p^0+E_u^{(+)}\right)\left(p^0+E_u^{(-)}\right)\left(p^0+E_{<}^{(+)}\right)\left(p^0+E_{<}^{(-)}\right)\big|_{\mathbf{p}\mapsto-\mathbf{p}} \notag \displaybreak[0]\\
&=\left[p^0-E_{<}^{(-)}(-d^{\mu\nu\dots},-H^{\mu\nu\dots})\right]\left[p^0-E_{<}^{(+)}(-d^{\mu\nu\dots},-H^{\mu\nu\dots})\right] \notag \displaybreak[0]\\
&\phantom{{}={}}\times\left(p^0-E_v^{(-)}\right)\left(p^0-E_v^{(+)}\right)\,.
\end{align}
\end{subequations}
The result involves the two possible antiparticle energies $E_v^{(\pm)}$ and the negative-energy values with the
signs of the $C$-odd coefficients $d^{\mu\nu}$ and $H^{\mu\nu}$ reversed. Based on this decomposition of the denominator,
the imaginary part of the forward scattering amplitude can be cast into the form
\begin{align}
\label{eq:forward-scattering-amplitude-positron}
2\mathrm{Im}(\overline{\mathcal{M}})&=\int \frac{\mathrm{d}^3p}{(2\pi)^32E_v^{(+)}}\delta^{(4)}(k_1+p_1-p)e^2\overline{v}^{(\alpha)}(p_1)\gamma^{\nu} \notag \displaybreak[0]\\
&\phantom{{}={}\int \frac{\mathrm{d}^3p}{(2\pi)^32E_v^{(+)}}}\hspace{-1cm}\times\left[-\mathscr{C}\left(E_v^{(-)},E^{(+)}_<(-d^{\mu\nu\dots},-H^{\mu\nu\dots}),E^{(-)}_<(-d^{\mu\nu\dots},-H^{\mu\nu\dots})\right)\right]\Big|_{p^0=E_v^{(+)}} \notag \displaybreak[0]\\
&\phantom{{}={}\int \frac{\mathrm{d}^3p}{(2\pi)^32E_v^{(+)}}}\hspace{-1cm}\times\left(\widehat{\xi}^{\,\mu}\gamma_{\mu}+\widehat{\Xi}+\widehat{\Upsilon}+\widehat{\zeta}^{\,\mu}\gamma_5\gamma_{\mu}+\widehat{\psi}^{\,\mu\nu}\sigma_{\mu\nu}\right)\Big|{\substack{p^0=-E_v^{(+)} \\ \hspace{-0.3cm}\mathbf{p}\mapsto -\mathbf{p}}} \notag \displaybreak[0]\\
&\phantom{{}={}\int \frac{\mathrm{d}^3p}{(2\pi)^32E_v^{(+)}}}\hspace{-1cm}\times\gamma^{\mu}v^{(\alpha)}(p_1)\varepsilon_{\mu}^{(\lambda)}(k_1)\overline{\varepsilon}_{\nu}^{(\lambda)}(k_1)\,,
\end{align}
with the scalar function $\mathscr{C}$ defined in \eqref{eq:scalar-function}. The remainder of the antifermion
propagator $S(-p)$ contains the latter scalar function and the matrix structure of the propagator, which is
evaluated at the antiparticle energy $E_v^{(+)}$ and with all spatial momentum components replaced by
their counterparts with opposite sign. The optical theorem says that \eqref{eq:forward-scattering-amplitude-positron}
is related to the cross section $\widetilde{\sigma}\equiv\sigma(\mathrm{e^+}\upgamma\rightarrow\mathrm{e^+})$,
with the amplitude $\widetilde{\mathcal{M}}\equiv \mathcal{M}(\mathrm{e^+}\upgamma\rightarrow \mathrm{e^+})$,
given as
\begin{align}
\label{eq:cross-section-positron}
\widetilde{\sigma}&=\int \frac{\mathrm{d}^3p}{(2\pi)^32E_v^{(+)}}\,\delta^{(4)}(k_1+p_2-p)|\widetilde{\mathcal{M}}|^2 \notag \displaybreak[0]\\
&=\int \frac{\mathrm{d}^3p}{(2\pi)^32E_v^{(+)}}\,\delta^{(4)}(k_1+p_1-p) \notag \displaybreak[0]\\
&\phantom{{}={}\int \frac{\mathrm{d}^3p}{(2\pi)^32E_v^{(+)}}}\times \mathrm{i}e\overline{v}^{(\alpha)}(p_1)\gamma^{\mu}v^{(1)}(p)\varepsilon^{(\lambda)}_{\mu}(k_1)\left(\mathrm{i}e\overline{v}^{(\alpha)}(p_1)\gamma^{\nu}v^{(1)}(p)\varepsilon_{\nu}^{(\lambda)}(k_1)\right)^{\dagger} \notag\displaybreak[0] \\
&=\int \frac{\mathrm{d}^3p}{(2\pi)^32E_v^{(+)}}\,\delta^{(4)}(k_1+p_1-p) \notag \displaybreak[0]\\
&\phantom{{}={}\int \frac{\mathrm{d}^3p}{(2\pi)^32E_v^{(+)}}}\times e^2\overline{v}^{(\alpha)}(p_1)\gamma^{\mu}\Big[v^{(1)}(p)\overline{v}^{(1)}(p)\Big]\gamma^{\nu} v^{(\alpha)}(p_1)\varepsilon_{\mu}^{(\lambda)}(k_1)\overline{\varepsilon}_{\nu}^{(\lambda)}(k_1)\,.
\end{align}
The calculation can be carried out one-to-one for the antiparticle energy $E_v^{(-)}$. With these results at
hand, we are able to simply read the spinor matrices $v^{(\alpha)}\overline{v}^{(\alpha)}$ off by comparing
Eqs.~(\ref{eq:forward-scattering-amplitude-positron}) and (\ref{eq:cross-section-positron}) directly to each other.

It is possible to treat the frameworks of the pseudovector coefficients $\hat{\mathcal{A}}^{\mu}$ and the two-tensor
coefficients $\hat{\mathcal{T}}^{\mu\nu}$ in one go. The spinor matrices for particles and antiparticles finally read as follows:
\begin{subequations}
\label{eq:spinor-matrices-general}
\begin{align}
u^{(1,2)}\overline{u}^{(1,2)}|_{E_u^{(\pm)}}&=\mathscr{C}(E_u^{(\mp)},E_<^{(+)},E_<^{(-)})|_{p^0=E_u^{(\pm)}} \notag \\
&\phantom{{}={}}\times\left(\widehat{\xi}^{\,\mu}_{\hat{\mathcal{A}},\hat{\mathcal{T}}}\gamma_{\mu}+\widehat{\Xi}_{\hat{\mathcal{A}},\hat{\mathcal{T}}}+\widehat{\Upsilon}_{\hat{\mathcal{A}},\hat{\mathcal{T}}}+\widehat{\zeta}^{\,\mu}_{\hat{\mathcal{A}},\hat{\mathcal{T}}}\gamma_5\gamma_{\mu}+\widehat{\psi}^{\,\mu\nu}_{\hat{\mathcal{A}},\hat{\mathcal{T}}}\sigma_{\mu\nu}\right)\Big|_{p^0=E_u^{(\pm)}}\,, \displaybreak[0]\\[2ex]
v^{(1,2)}\overline{v}^{(1,2)}|_{E_v^{(\pm)}}&=-\mathscr{C}\left(E_v^{(\mp)},E_<^{(+)}(-d^{\mu\nu\dots},-H^{\mu\nu\dots}),E_<^{(-)}(-d^{\mu\nu\dots},-H^{\mu\nu\dots})\right)\Big|_{p^0=E_v^{(\pm)}} \notag \\
&\phantom{{}={}}\times\left(\widehat{\xi}^{\,\mu}_{\hat{\mathcal{A}},\hat{\mathcal{T}}}\gamma_{\mu}+\widehat{\Xi}_{\hat{\mathcal{A}},\hat{\mathcal{T}}}+\widehat{\Upsilon}_{\hat{\mathcal{A}},\hat{\mathcal{T}}}+\widehat{\zeta}^{\,\mu}_{\hat{\mathcal{A}},\hat{\mathcal{T}}}\gamma_5\gamma_{\mu}+\widehat{\psi}^{\,\mu\nu}_{\hat{\mathcal{A}},\hat{\mathcal{T}}}\sigma_{\mu\nu}\right)\Big|{\substack{p^0=-E_v^{(\pm)} \\ \hspace{-0.3cm}\mathbf{p}\mapsto -\mathbf{p}}}\,,
\end{align}
\end{subequations}
with the scalar function $\mathscr{C}$ of \eqref{eq:scalar-function}. The matrix coefficients have to be taken
accordingly either from the propagator for $\hat{\mathcal{A}}^{\mu}$ or $\hat{\mathcal{T}}^{\mu\nu}$. Several remarks are in order. First, since these
expressions result from the exact propagators given in \secref{sec:propagator-pseudovector}, they are valid at all orders
in Lorentz violation. Second, they hold for on-shell particles, which is why all $p^0$ must be replaced by the appropriate
energies $E_u^{(\pm)}$ for particles and $E_v^{(\pm)}$ for antiparticles that are associated to the spinors.
Third, the results, as they stand, are only valid when there are no controlling coefficients that produce additional
time derivatives. How to treat such cases will be described in the latter \secref{sec:additional-time-derivatives}.
Fourth, for vanishing Lorentz violation, $\mathscr{C}\mapsto 1/(p^2-m_{\psi}^2)$, and looking at the propagator coefficients
of Eqs.~(\ref{eq:propagator-pseudovector}), (\ref{eq:propagator-two-tensor}) shows that the standard expressions of
\eqref{eq:spinor-matrices-standard} are recovered when Lorentz invariance is reestablished. Fifth, for the expressions
of \eqref{eq:spinor-matrices-general} we checked that
\begin{subequations}
\begin{align}
S^{-1}(p)\left(u^{(1,2)}\overline{u}^{(1,2)}|_{E_u^{(\pm)}}\right)&=0\,, \\[2ex]
S^{-1}(-p)\left(v^{(1,2)}\overline{v}^{(1,2)}|_{E_v^{(\pm)}}\right)&=0\,,
\end{align}
\end{subequations}
with the modified Dirac operator of \eqref{eq:dirac-operator}. Sixth, since the spinors span spinor space and they are
orthogonal to each other according to the orthogonality conditions of \secref{eq:orthogonality-relations}, adding up
all four spinor matrices, leads to the completeness relation
\begin{equation}
\label{eq:completeness-relation}
\sum_{\alpha=1,2} \left(\frac{1}{2E_u^{(\alpha)}}u^{(\alpha)}\overline{u}^{(\alpha)}|_{E_u^{(\alpha)}}+\frac{1}{2E_v^{(\alpha)}(-\mathbf{p})}v^{(\alpha)}(-\mathbf{p})\overline{v}^{(\alpha)}(-\mathbf{p})|_{E_v^{(\alpha)}}\right)\gamma^0=\mathds{1}_4\,.
\end{equation}
The latter is directly connected to Eq.~(23) in \cite{Kostelecky:2000mm} and it was verified to be valid at all orders
in Lorentz violation for the relations of \eqref{eq:spinor-matrices-general}. Furthermore, we were able to demonstrate
the validity of Eq.~(23) in \cite{Kostelecky:2000mm}, at first order in Lorentz violation, employing the spinor results
directly. Finally, note that for
the ``isotropic'' case of \secref{sec:isotropic-operator}, the $b$ and $d$ coefficients have to be treated separately
from the $g$ coefficients, as the first are comprised by $\hat{\mathcal{A}}^{\mu}$ and the latter are contained in
$\hat{\mathcal{T}}^{\mu\nu}$. Besides the general expressions stated above, several special results at first order in
Lorentz violation will be given in \appref{sec:special-spinor-matrices} to make the behavior of the spinor matrices more
transparent. We will not give explicit results that are exact in Lorentz violation as they do not provide any further
insight. Once a specific result is needed for practical purposes, it should be possible
to derive it from the general formulas of \eqref{eq:spinor-matrices-general}, with the help of computer algebra.

At this point, we will make a couple of final statements on certain conditions for the validity of the previous results.
For the pseudovector operator $\hat{\mathcal{A}}^{\mu}$, the spinors, the spinor matrices of \eqref{eq:spinor-matrices-general},
and the completeness relation of \eqref{eq:completeness-relation} are valid for controlling coefficients chosen such
that the following conditions are fulfilled:
\begin{subequations}
\begin{align}
0&\leq \hat{\mathcal{A}}^{\mu}\,, \displaybreak[0]\\[2ex]
0&\leq \hat{\mathcal{A}}^{\mu}(d^{\mu\nu\dots}\mapsto -d^{\mu\nu\dots})\,, \displaybreak[0]\\[2ex]
0&\leq \hat{\mathcal{A}}^{\mu}(d^{\mu\nu\dots}\mapsto -d^{\mu\nu\dots},\mathbf{p}\mapsto -\mathbf{p})\,.
\end{align}
\end{subequations}
The first of those ensures that the particle energies and spinors are numbered such as indicated throughout the paper.
The second grants the same for the antiparticle energies and spinors. The third applies to the completeness relation
and makes certain that the numbering of the antiparticle energies and spinors stays consistent. There exist controlling
coefficients and momentum components such that all three conditions are satisfied. For the two-tensor operator, there are
similar three conditions:
\begin{subequations}
\begin{align}
0&\leq\hat{\mathcal{T}}^{\mu\nu}\,, \displaybreak[0]\\[2ex]
0&\leq\hat{\mathcal{T}}^{\mu\nu}(H^{\mu\nu\dots}\mapsto -H^{\mu\nu\dots})\,, \displaybreak[0]\\[2ex]
0&\leq\hat{\mathcal{T}}^{\mu\nu}(H^{\mu\nu\dots}\mapsto -H^{\mu\nu\dots},\mathbf{p}\mapsto -\mathbf{p})\,.
\end{align}
\end{subequations}
The reasons for choosing these conditions are analog to the reasons outlined for $\hat{\mathcal{A}}^{\mu}$. However,
$\hat{\mathcal{T}}^{\mu\nu}$ differs from $\hat{\mathcal{A}}^{\mu}$ crucially. Although each of the conditions above can be fulfilled for
$\hat{\mathcal{A}}^{\mu}$, this is not possible for $\hat{\mathcal{T}}^{\mu\nu}$. Either the first two are true, but not the third or
vice versa. The reason for this is that the $g$ coefficients are contracted with one additional power of the momentum
compared to the $H$ coefficients. So, once the first two conditions are valid, changing the signs of the momentum
components obscures the third condition. This does not occur for the pseudovector $\hat{A}^{\mu}$, since it is the
$d$ coefficients that are contracted with one additional four-momentum relative to the $b$ coefficients. Hence, once the
first two conditions are valid for $\hat{\mathcal{A}}^{\mu}$, the third is not compromised by changing the signs of the
momentum components. Now let us assume that the first two conditions for $\hat{T}^{\mu\nu}$ hold. The consequence is
then that the antiparticle energies have to be switched in the completeness relation of \eqref{eq:completeness-relation}.

\section{Additional time derivatives}
\label{sec:additional-time-derivatives}
\setcounter{equation}{0}

The results established in the previous section with the help of the optical theorem are only valid in case there are
not any controlling coefficients that introduce additional time derivatives into the Lagrangian. Such time derivatives
lead to an unconventional time evolution of the physical states \cite{Colladay:2001wk}. Furthermore, for the nonminimal
SME they increase the degree of the polynomial in $p^0$ that follows from the determinant of the Dirac operator. First,
this may lead to additional spurious dispersion relations that do not have Lorentz-invariant equivalents. Second, the
structure of the denominator $\Delta$ in propagators is modified drastically, which renders the previous proof based
on the optical theorem invalid \cite{Schreck:2013kja,Schreck:2014qka}.

\subsection{Minimal fermion sector}

One possibility of remedying this behavior at least for the minimal SME was outlined in \cite{Colladay:2001wk}. The
authors of the latter paper suggest introducing a new spinor $\chi$, which is linked to the former spinor $\psi$ via
a transformation with an invertible matrix $A$: $\psi=A\chi$. The matrix $A$ shall be constructed such that
$A^{\dagger}\gamma^0\Gamma^0 A=\mathds{1}_4$. Since $\Gamma^0$ is linked to additional time derivatives in the
minimal fermion sector all such time derivatives can be removed in this way.

The procedure will be exemplified within the minimal fermion sector with a nonzero coefficient $d^{(4)00}$. The matrix
$A$ is found by solving the matrix equation $A^{\dagger}\gamma^0(\gamma^0+d^{(4)00}\gamma^5\gamma^0)A=\mathds{1}_4$.
Making the \textit{Ansatz} of a diagonal $A$ with real coefficients allows for a convenient solution of the system
and it gives:
\begin{equation}
A=\mathrm{diag}\left(\frac{1}{\sqrt{1+d^{(4)00}}},\frac{1}{\sqrt{1+d^{(4)00}}},\frac{1}{\sqrt{1-d^{(4)00}}},\frac{1}{\sqrt{1-d^{(4)00}}}\right)\,.
\end{equation}
Introducing the new spinor $\chi$ into the Dirac operator leads to
\begin{equation}
\mathcal{L}=\frac{1}{2}\overline{\chi}S'^{-1}\chi+\text{H.c.}\,,\quad S'^{-1}=\gamma^0A^{\dagger}\gamma^0(\gamma^{\mu}\mathrm{i}\partial_{\mu}-m_{\psi}\mathds{1}_4+\hat{\mathcal{Q}})A\,.
\end{equation}
Note that the dispersion relations are not modified by this transformation but the propagator is. The new propagator
$S'$ can be shown to be of the following shape where the purely timelike four-vector $\lambda^{\mu}=(1,0,0,0)^{\mu}$
is introduced for convenience:
\begin{subequations}
\label{eq:propagator-isotropic-d4}
\begin{align}
S'&=\frac{1}{\Delta}\left(\widehat{\xi}'_{\mu}\gamma^{\mu}+\widehat{\Xi}'\mathds{1}_4+\widehat{\Upsilon}'\gamma^5+\widehat{\zeta}'_{\mu}\gamma^5\gamma^{\mu}+\widehat{\psi}'_{\mu\nu}\sigma^{\mu\nu}\right)\,, \\[2ex]
\widehat{\Xi}'&=(A_{11}A_{33})^{-1}\widehat{\Xi}\,, \displaybreak[0]\\[2ex]
\widehat{\Upsilon}'&=(A_{11}A_{33})^{-1}\widehat{\Upsilon}\,, \displaybreak[0]\\[2ex]
\widehat{\xi}'^{\mu}&=\widehat{\xi}^{\mu}+\frac{1}{27}(\alpha\lambda^{\mu}+\beta p^{\mu})\,, \displaybreak[0]\\[2ex] \widehat{\zeta}'^{\mu}&=\widehat{\zeta}^{\mu}+\frac{1}{9}(\gamma\lambda^{\mu}+\delta p^{\mu})\,, \displaybreak[0]\\[2ex]
\widehat{\psi}'^{\mu\nu}&=(A_{11}A_{33})^{-1}\widehat{\psi}^{\mu\nu}\,,
\end{align}
with the helpful quantities
\begin{align}
\alpha&=4p^0(d^{(4)00})^2\left\{8(d^{(4)00})^2(p\cdot\lambda)^2+\left[9+(d^{(4)00})^2\right]p^2+9m_{\psi}^2\right\}\,, \displaybreak[0]\\[2ex]
\beta&=-(d^{(4)00})^2\left\{8\left[9+(d^{(4)00})^2\right](p\cdot\lambda)^2-\left[9-(d^{(4)00})^2\right]p^2+9m_{\psi}^2\right\}\,, \displaybreak[0]\\[2ex]
\gamma&=8p^0(d^{(4)00})^3\left[4(p\cdot\lambda)^2-p^2\right]\,, \displaybreak[0]\\[2ex]
\delta&=-d^{(4)00}\left\{16(d^{(4)00})^2(p\cdot\lambda)^2+\left[9-(d^{(4)00})^2\right]p^2-9m_{\psi}^2\right\}\,.
\end{align}
\end{subequations}
Hence, the scalar, pseudoscalar, and the tensor parameters are just multiplied with a global prefactor, which corresponds
to the inverse product of two matrix components. The modification of the vector and pseudovector parameters is more
involved, though. As expected, the denominator $\Delta$ remains unchanged since it corresponds to the determinant of the
Dirac operator modulo a global prefactor.

As the Dirac operator has changed the spinors will change as well. The new spinors are given by $\chi=A^{-1}\psi$ where
the spinors $\psi$ for particles at first order in Lorentz violation are stated in \eqref{eq:particle-spinors-b0} and for
antiparticles they are obtained from \eqref{eq:antiparticle-spinors-pseudovector}. Hence, each of the already known spinors
just has to be multiplied with the inverse of $A$, which is also a diagonal matrix. Note that the normalization must be
adapted as well. Based on these spinors, the spinor matrices at first order in Lorentz violation are derived:
\begin{subequations}
\begin{align}
u^{(1,2)}\overline{u}^{(1,2)}&=\upsilon_{\pm}\left(\xi^{\mu}\gamma_{\mu}+\Xi\mathds{1}_4\pm \zeta^{\mu}\gamma_5\gamma_{\mu}\pm \psi^{\mu\nu}\sigma_{\mu\nu}\right)\big|_{p^0=E_0}\,, \displaybreak[0]\\[2ex]
v^{(1,2)}\overline{v}^{(1,2)}&=\upsilon_{\pm}\left(\xi^{\mu}\gamma_{\mu}-\Xi\mathds{1}_4\pm \zeta^{\mu}\gamma_5\gamma_{\mu}\mp \psi^{\mu\nu}\sigma_{\mu\nu}\right)\big|_{p^0=E_0}\,, \displaybreak[0]\\[2ex]
\xi^{\mu}&=\frac{p^{\mu}}{2}\,, \displaybreak[0]\\[2ex]
\zeta^0&=\frac{|\mathbf{p}|}{2}\,,\quad \boldsymbol{\zeta}=\frac{E_0}{|\mathbf{p}|}\frac{\mathbf{p}}{2}\,, \displaybreak[0]\\[2ex]
\psi^{0i}&=0\,,\quad \psi^{ij}=\frac{m_{\psi}}{4|\mathbf{p}|}\varepsilon^{ijk}p^k\,, \displaybreak[0]\\[2ex]
\Xi&=\frac{m_{\psi}}{2}\,, \displaybreak[0]\\[2ex]
\upsilon_{\pm}&=1\pm\frac{4}{3}\frac{|\mathbf{p}|}{E_0}d^{(4)00}\,.
\end{align}
\end{subequations}
Using the results for the modified propagator of \eqref{eq:propagator-isotropic-d4} and the previously stated spinor
matrices allows for demonstrating the validity of the optical theorem at first order in Lorentz violation. A reasonable
assumption is that the optical theorem is valid at all orders in Lorentz violation,
cf.~\cite{Klinkhamer:2010zs,Schreck:2011ai,Schreck:2013gma,Schreck:2013kja,Maniatis:2014xja,Schreck:2014qka}.
Then the spinor matrices are given by \eqref{eq:spinor-matrices-general} based on the modified propagator quantities
stated in \eqref{eq:propagator-isotropic-d4}.

\subsection{Nonminimal fermion sector}

In the nonminimal SME, there is an infinite number of component coefficients that are contracted with additional time
derivatives. This leads to problems similar to
the ones discussed in the previous section. One of the simplest examples within the scope of the paper is the operator
$b^{(5)000}(\mathrm{i}\partial^0)^2$ that is contracted with two time derivatives. Such nonminimal frameworks can have
spurious dispersion relations that do not have equivalents in the standard theory:
\begin{subequations}
\label{eq:dispersion-spurious-b5000}
\begin{align}
(p^0)^{(\pm)}&=\frac{1}{b^{(5)000}}\pm |\mathbf{p}|-\frac{1}{2}(2\mathbf{p}^2+m_{\psi}^2)b^{(5)000}+\dots\,, \\[2ex]
(p^0)^{(\pm)}&=-\frac{1}{b^{(5)000}}\pm |\mathbf{p}|+\frac{1}{2}(2\mathbf{p}^2+m_{\psi}^2)b^{(5)000}+\dots\,.
\end{align}
\end{subequations}
Dispersion relations like these could be interpreted as Planck-scale effects. However, since the realm of applicability of
the SME is expected to be far below the Planck energy (see also the issues connected to highly boosted observer
frames in \cite{Kostelecky:2000mm}) such dispersion relations are usually discarded in any analysis. Note that the
validity of the optical theorem is obscured by the occurrence of spurious dispersion relations
\cite{Schreck:2013kja,Schreck:2014qka}.

In contrast to the minimal fermion sector, contributions with additional time derivatives in the nonminimal sector do
not modify the matrix structure in spinor space, which is why the additional time derivatives cannot be simply removed
with a matrix transformation along the lines of the previous section. One method of dealing with such terms at least
at first order in Lorentz violation was introduced in \cite{Schreck:2013kja,Schreck:2014qka}. It amounts to replacing
all zeroth four-momentum components in the Lorentz-violating terms of the Dirac operator by the standard dispersion
relation $p^0=E_0=\sqrt{\mathbf{p}^2+m_{\psi}^2}$. This does not modify the physical dispersion relations at first order
but it removes the spurious dispersion relations given by \eqref{eq:dispersion-spurious-b5000}. As the current
framework is isotropic the spinor solutions of \eqref{eq:particle-spinors-b0} must be taken with all $p^0$ in the
Lorentz-violating terms replaced by $E_0$, which was already indicated in \secref{sec:obtaining-spinors} anyhow.
Additionally, the same should be carried out in the propagator where it has to be kept in mind that this modified
propagator can only be used in the proof of the optical theorem. In general, a propagator describes virtual (off-shell)
particles, which is why $p^0$ does not satisfy the dispersion relation in this context. Based on that simple
modification, the validity of the optical theorem is again established at first order in Lorentz violation. Hence,
the spinor matrices are obtained from \eqref{eq:spinor-matrices-general} with $p^0=E_0$ in the Lorentz-violating
contributions. The corresponding explicit results will be stated in \appref{sec:special-spinor-matrices}.

\section{Conclusions}
\label{sec:conclusions}
\setcounter{equation}{0}

In the current paper, we have examined the fermion sector of the Standard-Model Extension for the spin-nondegenerate Lorentz-violating
operators. This concerned the $b$, $d$, $H$, and $g$ coefficients where the first two are contained in the pseudovector
$\hat{\mathcal{A}}^{\mu}$ and the latter two are contained in the two-tensor $\hat{\mathcal{T}}^{\mu\nu}$. We obtained the
modified propagators for both $\hat{\mathcal{A}}^{\mu}$ and $\hat{\mathcal{T}}^{\mu\nu}$. These results are exact in
Lorentz violation and they are valid for coefficients of both the minimal and the nonminimal SME.

Also, the dispersion relations and solutions of the modified Dirac equation, i.e., the spinors for particles and antiparticles
were computed at first order in Lorentz violation for particular families of both minimal and nonminimal coefficients
in $\hat{\mathcal{A}}^{\mu}$ and $\hat{\mathcal{T}}^{\mu\nu}$. With the optical theorem, the propagators and the spinors were checked
to be consistent with each other at first order in Lorentz violation. Under the assumption that the optical
theorem is valid at tree-level at all orders in Lorentz violation, which is reasonable based on earlier investigations, the
spinor matrices $u\overline{u}$ and $v\overline{v}$ were extracted from the propagator at all orders in Lorentz violation.

The expressions obtained will prove useful for both future theoretical and phenomenological studies. First, the spinor solutions
of the Dirac equation and especially their nonrelativistic limits may be employed for phenomenology in fermion systems where the
electron spin is essential and cannot be neglected. Second, the spinor matrices and propagators are needed for
computations in high-energy physics that are carried out in the context of quantum field theory. Such investigations are currently
in progress and the findings will be reported elsewhere.

\section{Acknowledgments}

It is a pleasure to thank V.A.~Kosteleck\'{y} and M.M. Ferreira Jr for several helpful suggestions. This work was partially funded
by the Brazilian foundation FAPEMA.

\newpage
\begin{appendix}
\numberwithin{equation}{section}

\section{Exact spinors for special frameworks}
\label{sec:exact-spinors}
\setcounter{equation}{0}

In this part of the appendix, we demonstrate how to obtain exact solutions of the modified Dirac equation. Doing this with computer algebra is
unproblematic as long as there is a single nonzero controlling coefficient only. However, the complexity of both the modified dispersion relations
and the spinor solutions rises drastically with an increasing number of coefficients. This is the main reason why most of the computations were
restricted to first order in Lorentz violation, which allows to obtaining the energies and spinors for large sets of coefficients --- even including
nonminimal ones. For brevity, we will study isotropic cases here only.

\subsection{Isotropic $\boldsymbol{b}$ coefficients}
\label{sec:exact-spinors-isotropic-b}

The first framework shall be characterized by a nonzero $b^{(3)0}$. The particle dispersion relations are obtained directly from the determinant of
the Dirac operator:
\begin{equation}
E_u^{(\pm)}=\sqrt{\mathbf{p}^2+m_{\psi}^2\mp 2b^{(3)0}|\mathbf{p}|+(b^{(3)0})^2}\,.
\end{equation}
There are two distinct ones, as expected, and they differ from each other at first order in Lorentz violation. Plugging these energies into the Dirac
equation and solving it leads to the particle spinor solutions:
\begin{equation}
u^{(1,2)}|_{E_u^{(\pm)}}(\mathbf{p})=N_u^{(1,2)}\begin{pmatrix}
(p_3\pm|\mathbf{p}|)(E_u^{(\pm)}-p_3+b^{(3)0})-(p_1^2+p_2^2) \\
(p_1+\mathrm{i}p_2)(E_u^{(\pm)}\mp |\mathbf{p}|+b^{(3)0}) \\
m_{\psi}(p_3\pm |\mathbf{p}|) \\
m_{\psi}(p_1+\mathrm{i}p_2) \\
\end{pmatrix}\,.
\end{equation}
The antiparticle spinors are connected to the particle spinors according to \eqref{eq:antiparticle-spinors-pseudovector}.
The normalizations for the particle spinors are chosen such that $(u^{(\alpha)})^{\dagger}u^{(\beta)}=2E_u^{(\alpha)}\delta_{\alpha\beta}$
and in analogy for the antiparticle spinors. This also requires that spinors for different energies are orthogonal to each other, which is
granted by the hermiticity of the Dirac operator. Furthermore, such a normalization allows for a convenient check of the optical theorem,
which we saw in \secref{sec:spinor-matrices-optical-theorem}. For this particular case the normalizations read as follows:
\begin{equation}
N_u^{(1,2)}=\frac{1}{\sqrt{2|\mathbf{p}|(|\mathbf{p}|\pm p_3)(E_u^{(\pm)}\mp |\mathbf{p}|+b^{(3)0})}}\,,\quad N_v^{(1,2)}=N_u^{(2,1)}\,.
\end{equation}


\subsection{Isotropic $\boldsymbol{g}$ coefficients}

In contrast to the $H$ coefficients, there is an isotropic case for the $g$ coefficients. It is characterized by a totally antisymmetric choice
of nonzero coefficients: $g^{(4)ijk}\equiv \varepsilon^{ijk}g_1$. where $i$, $j$, and $k$ are spatial indices.
There are again two dispersion relations that differ from each other at first order in Lorentz violation:
\begin{equation}
E_u^{(\pm)}=\sqrt{(1+g_1^2)\mathbf{p}^2\pm 2g_1m_{\psi}|\mathbf{p}|+m_{\psi}^2}\,.
\end{equation}
Note that the $g$ coefficients are dimensionless. The particle spinor solutions can be obtained such as before:
\begin{equation}
u^{(1,2)}|_{E_u^{(\pm)}}(\mathbf{p})=N_u^{(1,2)}\begin{pmatrix}
-\left[p_1^2+p_2^2\pm(p_3-E_u^{(\pm)})(|\mathbf{p}|\pm p_3)\right] \\
(p_1+\mathrm{i}p_2)(E_u^{(\pm)}\mp |\mathbf{p}|) \\
(p_3\pm|\mathbf{p}|)(m_{\psi}\pm g_1|\mathbf{p}|) \\
(p_1+\mathrm{i}p_2)(m_{\psi}\pm g_1|\mathbf{p}|) \\
\end{pmatrix}\,,
\end{equation}
and the antiparticle spinor solutions are connected to these based on \eqref{eq:antiparticle-spinors-two-tensor}.
The normalization factors are simply expressed in terms of the particle energies:
\begin{equation}
N_u^{(1,2)}=\frac{1}{\sqrt{2|\mathbf{p}|(|\mathbf{p}|\pm p_3)(E_u^{(\pm)}\mp |\mathbf{p}|)}}\,,\quad N_v^{(1,2)}=N_u^{(2,1)}\,.
\end{equation}
The exact spinors in both subsections were checked to correspond to the results in \secref{sec:isotropic-operator}
at first order in Lorentz violation.

\section{Spinors for spin-degenerate operators}
\label{sec:results-spin-degenerate}
\setcounter{equation}{0}

For completeness, we state the spinors, normalization factors, and the propagator for the spin-degenerate cases that are
encoded in the scalar operator $\hat{\mathcal{S}}$ and the vector operator $\hat{\mathcal{V}}^{\mu}$. The Dirac operator can be
diagonalized with the matrix $U=V\cdot W$ where $V$ and $W$ are given by \eqref{eq:transformation-matrices}. In these matrices all
$E_0$ have to be replaced by the exact dispersion relation $E_u$ and all occurrences of $m_{\psi}$ by $m_{\psi}-\hat{\mathcal{S}}$
and $p^{\mu}$ by $(p+\hat{\mathcal{V}})^{\mu}$ \cite{Kostelecky:2013rta}. In the chiral representation the particle spinors
at all orders in Lorentz violation can be cast into the following form:
\begin{subequations}
\begin{align}
\frac{1}{N_u}u^{(1)}|_{E_u}&=(m_{\psi}-\hat{\mathcal{S}})\begin{pmatrix}
0 \\
1 \\
0 \\
1 \\
\end{pmatrix}+\begin{pmatrix}
-(p+\hat{\mathcal{V}})^1+\mathrm{i}(p+\hat{\mathcal{V}})^2 \\
E_u+\hat{\mathcal{V}}^0+(p+\hat{\mathcal{V}})^3 \\
(p+\hat{\mathcal{V}})^1-\mathrm{i}(p+\hat{\mathcal{V}})^2 \\
E_u+\hat{\mathcal{V}}^0-(p+\hat{\mathcal{V}})^3 \\
\end{pmatrix}\,, \\[2ex]
\frac{1}{N_u}u^{(2)}|_{E_u}&=(m_{\psi}-\hat{\mathcal{S}})\begin{pmatrix}
1 \\
0 \\
1 \\
0 \\
\end{pmatrix}+\begin{pmatrix}
E_u+\hat{\mathcal{V}}^0-(p+\hat{\mathcal{V}})^3 \\
-(p+\hat{\mathcal{V}})^1-\mathrm{i}(p+\hat{\mathcal{V}})^2 \\
E_u+\hat{\mathcal{V}}^0+(p+\hat{\mathcal{V}})^3 \\
(p+\hat{\mathcal{V}})^1+\mathrm{i}(p+\hat{\mathcal{V}})^2 \\
\end{pmatrix}\,,
\end{align}
\end{subequations}
where $E_u$ are the exact particle energies.
The antiparticle spinors are obtained from the particle spinors by charge conjugation (cf.~\secref{sec:remaining-spinors-pseudovector}):
\begin{equation}
v^{(1,2)}|_{E_v}=\begin{pmatrix}
u^{(2,1)}_4 \\
-u^{(2,1)}_3 \\
-u^{(2,1)}_2 \\
u^{(2,1)}_1 \\
\end{pmatrix}^{*}(-a^{\mu\nu\dots},-e^{\mu\nu\dots})\,,\quad E_v=E_u(-a^{\mu\nu\dots},-e^{\mu\nu\dots})\,.
\end{equation}
with the signs of the $C$-odd $a$ and $e$ coefficients reversed (see Tab. P31 in \cite{Kostelecky:2008ts}). The normalization factors for particles
and antiparticles read
\begin{equation}
N_u=\sqrt{\frac{E_u}{(E_u+\hat{\mathcal{V}}^0+m_{\psi}-\hat{\mathcal{S}})^2+(\mathbf{p}+\boldsymbol{\hat{\mathcal{V}}})^2}}\,,\quad N_v=N_u(-a^{\mu\nu\dots},-e^{\mu\nu\dots})\,,
\end{equation}
where $\boldsymbol{\hat{\mathcal{V}}}$ is the spatial part of the vector operator $\hat{\mathcal{V}}^{\mu}$. Last but not least,
the propagator can be expressed as follows:
\begin{equation}
\mathrm{i}S=\frac{\mathrm{i}(\cancel{p}+\cancel{\hat{\mathcal{V}}}+(m_{\psi}-\hat{\mathcal{S}})\mathds{1}_4)}{(p+\hat{\mathcal{V}})^2-(m_{\psi}-\hat{\mathcal{S}})^2}\,.
\end{equation}

\section{Spinors for pseudoscalar operator $\boldsymbol{\hat{f}}$}

The spinors for the pseudoscalar operator $\hat{\mathcal{Q}}=\mathrm{i}\hat{f}\gamma^5$ with $\hat{f}=\hat{f}^{\nu}p_{\nu}$ shall be stated as well.
We do not consider the chiral mass term $\mathrm{i}\hat{m}_5\gamma^5$, which in principle also adds to the pseudoscalar operator but can be rotated
away by a chiral transformation in many cases \cite{Kostelecky:2013rta}. The operator $\hat{f}$ has the peculiarity that its contributions to the
dispersion relation are of quadratic order at least, i.e., $E_u=E_0+\hat{f}^2/(2E_0)$. However, the spinors contain first-order terms in Lorentz violation:
\begin{subequations}
\begin{align}
\frac{1}{N_u}u^{(1)}|_{E_u}&=\begin{pmatrix}
\sigma^3\phi_- \\
\phi_+ \\
\end{pmatrix}-\frac{\mathrm{i}\hat{f}}{2E_0}\begin{pmatrix}
\phi_+ \\
-\sigma^3\phi_- \\
\end{pmatrix}\,, \\[2ex]
\frac{1}{N_u}u^{(2)}|_{E_u}&=\begin{pmatrix}
-\mathrm{i}\sigma^2\phi_+^{*} \\
\sigma^1\phi_-^{*} \\
\end{pmatrix}-\frac{\mathrm{i}\hat{f}}{2E_0}\begin{pmatrix}
\sigma^1\phi_-^{*} \\
\mathrm{i}\sigma^2\phi_+^{*} \\
\end{pmatrix}\,, \\[2ex]
\phi_{\pm}&=\begin{pmatrix}
A_{\pm} \\
B \\
\end{pmatrix}\,, \\[2ex]
A_{\pm}&=E_0+m_{\psi}\pm p_3\,, \\[2ex]
B&=p_1+\mathrm{i}p_2\,,
\end{align}
\end{subequations}
with the Pauli matrices $\sigma^i$. The antiparticle spinors can be obtained as usual where the signs of the $C$-odd coefficients $f^{\mu}$ must be reversed
(see Tab. P31 in \cite{Kostelecky:2008ts}):
\begin{equation}
v^{(1,2)}|_{E_v}=\begin{pmatrix}
u^{(2,1)}_4 \\
-u^{(2,1)}_3 \\
-u^{(2,1)}_2 \\
u^{(2,1)}_1 \\
\end{pmatrix}^{*}(-f^{\mu\dots})\,,\quad E_v=E_u\,.
\end{equation}
Last but not least, the normalization factors of the spinors and antispinors read
\begin{equation}
N_u=\frac{1}{\sqrt{2(E_0+m_{\psi})[1+\hat{f}^2/(4E_0^2)]}}\,,\quad N_v=N_u(-f^{\mu\dots})\,,
\end{equation}
and the propagator is given by
\begin{equation}
\mathrm{i}S=\frac{\mathrm{i}(\cancel{p}+m_{\psi}\mathds{1}_4+\mathrm{i}\hat{f}\gamma^5)}{p^2-m_{\psi}^2-\hat{f}^2}\,.
\end{equation}
In the denominator the square of $\hat{f}$ appears again demonstrating that $\hat{f}$ contributes to the dispersion
relation quadratically. The latter result corresponds to Eq.~(6.49) in \cite{Schreck:2014qka}.

\section{Special results for spinor matrices}
\label{sec:special-spinor-matrices}
\setcounter{equation}{0}

Finally, in the current section some explicit results for spinor matrices $u^{(\alpha)}\overline{u}^{(\alpha)}$ for particles
and $v^{(\alpha)}\overline{v}^{(\alpha)}$ for antiparticles shall be presented. They are based on the general expressions
stated in \eqref{eq:spinor-matrices-general} and will be given for the purpose of illustration. Two cases for the
pseudovector operator $\hat{\mathcal{A}}^{\mu}$ and one case for the two-tensor operator $\hat{\mathcal{T}}^{\mu\nu}$
will be under consideration.

\subsection{Pseudovector operator}

The first case is characterized by isotropic Lorentz violation and it was examined in \secref{sec:isotropic-operator}.
The spinor matrices for particles and antiparticles are explicitly given by
\begin{subequations}
\label{eq:spinor-matrices-perturbative-isotropic}
\begin{align}
u^{(1,2)}\overline{u}^{(1,2)}|_{E_u^{(\pm)}}&=\xi^{\mu}_{\pm}\gamma_{\mu}+\Xi\mathds{1}_4+\zeta^{\mu}_{\pm}\gamma_5\gamma_{\mu}+\psi_{\pm}^{\mu\nu}\sigma_{\mu\nu}\big|_{p^0=E_u^{(\pm)}}\,, \displaybreak[0]\\[2ex]
v^{(1,2)}\overline{v}^{(1,2)}|_{E_v^{(\pm)}}&=\xi^{\mu}_{\mp}\gamma_{\mu}-\Xi\mathds{1}_4-\zeta^{\mu}_{\mp}\gamma_5\gamma_{\mu}-\psi_{\pm}^{\mu\nu}\sigma_{\mu\nu}\big|_{p^0=E_v^{(\pm)}}\,, \displaybreak[0]\\[2ex]
\xi^0_{\pm}&=\frac{p^0}{2}\,,\quad \boldsymbol{\xi}_{\pm}=\frac{\mathbf{p}}{2}\left(1\mp\frac{b^{(3)0}}{|\mathbf{p}|}\right)\,, \displaybreak[0]\\[2ex]
\zeta^0_{\pm}&=\frac{1}{2}(\pm|\mathbf{p}|-b^{(3)0})\,,\quad \boldsymbol{\zeta}_{\pm}=\pm\frac{p^0}{2|\mathbf{p}|}\mathbf{p}\,, \displaybreak[0]\\[2ex]
\psi^{0i}_{\pm}&=0\,,\quad \psi^{ij}_{\pm}=\pm\frac{m_{\psi}}{4|\mathbf{p}|}\varepsilon^{ijk}p^k\,, \displaybreak[0]\\[2ex]
\Xi&=\frac{m_{\psi}}{2}\,.
\end{align}
\end{subequations}
Due to isotropy, no momentum component is preferred in these expressions. Furthermore, all parameter functions decompose into
a vector or tensor part and a scalar part that depends on the magnitude of the momentum only. Note that the signs of $\Xi$ for
particles and antiparticles are opposite, as expected.

The second framework is characterized by a nonvanishing $\hat{\mathcal{A}}^3$, which makes it anisotropic,
cf.~\secref{sec:anisotropic-pseudovector-a3}. With the preferred purely spacelike direction $\lambda^{\mu}=(0,0,0,1)^{\mu}$,
the spinor matrices can be expressed conveniently:
\begin{subequations}
\begin{align}
u^{(1,2)}\overline{u}^{(1,2)}|_{E_u^{(\pm)}}&=\xi^{\mu}_{\pm}\gamma_{\mu}+\Xi_{\pm}\mathds{1}_4+\zeta^{\mu}_{\pm}\gamma_5\gamma_{\mu}+\psi_{\pm}^{\mu\nu}\sigma_{\mu\nu}\big|_{p^0=E_u^{(\pm)}}\,, \displaybreak[0]\\[2ex]
v^{(1,2)}\overline{v}^{(1,2)}|_{E_v^{(\pm)}}&=\xi^{\mu}_{\mp}\gamma_{\mu}-\Xi_{\mp}\mathds{1}_4-\zeta^{\mu}_{\mp}\gamma_5\gamma_{\mu}-\psi_{\pm}^{\mu\nu}\sigma_{\mu\nu}\big|{\substack{p^0=E_v^{(\pm)} \\ d^{\mu\nu}\mapsto -d^{\mu\nu}}}\,, \displaybreak[0]\\[2ex]
\xi^{\mu}_{\pm}&=\frac{p^{\mu}}{2}\mp \frac{\hat{\mathcal{A}}^3}{2S_3}(p\cdot\lambda)\lambda^{\mu}\,, \\[2ex]
\zeta^{\mu}_{\pm}&=\frac{1}{2}\left[\pm\left(\frac{(p\cdot\lambda)^2}{S_3}-S_3\right)-\hat{\mathcal{A}}^3\right]\lambda^{\mu}\pm \frac{p\cdot\lambda}{2S_3}p^{\mu}\,, \displaybreak[0]\\[2ex]
\psi_{\pm}^{0i}&=\pm\frac{m_{\psi}}{4S_3}\varepsilon^{ij3}p^j\,,\quad
\psi_{\pm}^{ij}=\mp\frac{m_{\psi}p^0}{4S_3}\varepsilon^{ijk}\lambda^k\,, \displaybreak[0]\\[2ex]
\Xi_{\pm}&=\frac{m_{\psi}}{2}\left(1\pm \frac{\hat{\mathcal{A}}^3}{S_3}\right)\,.
\end{align}
\end{subequations}
Since this case is anisotropic with the preferred direction pointing along the third spatial axis of the coordinate system
the third spatial momentum component is preferred. This manifests in expressions depending on $S_3=\sqrt{p_3^2+m_{\psi}^2}$
and the scalar product $p\cdot\lambda$. Furthermore, due to the preferred spatial axis, the quantity $\Xi$ becomes
nondegenerate for the particle and antiparticle spinors, respectively.

\subsection{Two-tensor operator}

As an example for the two-tensor operator $\hat{\mathcal{T}}^{\mu\nu}$, a nonzero $\hat{\mathcal{T}}^{01}$ is
considered, cf.~\secref{eq:first-spinor-pseudoscalar-mixed}. For convenience, we introduce the three-dimensional vectors
$\mathbf{p}^{(3)}\equiv (0,p_2,p_3)$ and $\widetilde{\mathbf{p}}^{(3)}=\varepsilon^{1ij}\mathbf{p}^{(3)j}=(0,p_3,-p_2)$.
The parameters of the spinor matrices can then be written in a handy form:
\begin{subequations}
\begin{align}
u^{(1,2)}\overline{u}^{(1,2)}|_{E_u^{(\pm)}}&=\xi^{\mu}_{\pm}\gamma_{\mu}+\Xi\mathds{1}_4+\zeta_{\pm}^{\mu}\gamma_5\gamma_{\mu}+\psi^{\mu\nu}_{\pm}\sigma_{\mu\nu}\big|_{p^0=E_u^{(\pm)}}\,, \displaybreak[0]\\[2ex]
v^{(1,2)}\overline{v}^{(1,2)}|_{E_v^{(\pm)}}&=\xi^{\mu}_{\mp}\gamma_{\mu}-\Xi\mathds{1}_4-\zeta_{\mp}^{\mu}\gamma_5\gamma_{\mu}+\psi^{\mu\nu}_{\mp}\sigma_{\mu\nu}\big|{\substack{p^0=E_v^{(\pm)} \\ h^{\mu\nu}\mapsto -h^{\mu\nu}}}\,, \displaybreak[0]\\[2ex]
\xi^{\mu}_{\pm}&=\frac{p^{\mu}}{2}\pm \frac{\hat{\mathcal{T}}^{01}}{2S_1}(0,\mathbf{p}^{(3)})^{\mu}\,, \displaybreak[0]\\[2ex]
\zeta^{\mu}_{\pm}&=\pm\frac{m_{\psi}}{2S_1}(0,\widetilde{\mathbf{p}}^{(3)})^{\mu}\,, \displaybreak[0]\\[2ex]
\psi^{01}_{\pm}&=\frac{1}{4}(\pm S_1+\hat{\mathcal{T}}^{01})\,,\quad \psi^{0i}_{\pm}=\mp \mathbf{p}^{(3)i}\frac{p_1}{4S_1}\,,\quad
\psi^{ij}_{\pm}=\pm \varepsilon^{ijk}\widetilde{\mathbf{p}}^{(3)k}\frac{p_0}{4S_1}\,, \displaybreak[0]\\[2ex]
\Xi&=\frac{m_{\psi}}{2}\,,
\end{align}
\end{subequations}
with $S_1$ from \eqref{eq:particle-energies-T01-T02}.
Note that it is possible to express the vectors in $\xi^{\mu}_{\pm}$ and $\zeta^{\mu}_{\pm}$ directly in terms of the two-tensor
operator under consideration, i.e.,
\begin{equation}
(0,\widetilde{\mathbf{p}}^{(3)})^{\mu}=-\frac{\widetilde{\hat{\mathcal{T}}}{}^{\mu\nu}p_{\nu}}{\hat{\mathcal{T}}^{01}}\,,\quad (0,\mathbf{p}^{(3)})^{\mu}=\frac{\varepsilon^{01\mu\nu}\widetilde{\hat{\mathcal{T}}}{}^{\phantom{\nu}\rho}_{\nu}p_{\rho}}{\hat{\mathcal{T}}^{01}}\,.
\end{equation}
This concludes the examples for the spinor matrices that shall be stated explicitly.

\end{appendix}

\newpage


\end{document}